\documentclass[amsmath,amssymb,aps,showpacs,twocolumn,floats,superscriptaddress,floatfix]{revtex4}

\usepackage{graphicx}
\usepackage{amssymb}
\usepackage{amsmath}
\usepackage{amssymb}
\usepackage{bbold}
\usepackage{dsfont}
\usepackage{hyperref}
\usepackage{footnote}
\usepackage{color}
\usepackage{multirow}
\usepackage{enumerate}
\usepackage{soul}
\usepackage[FIGTOPCAP]{subfigure}
\pdfoutput=1 
\newcommand{\matr}[1]{\mathbf{#1}}

\begin{document}
 
\title{Kerr enhanced optomechanical cooling in the unresolved sideband regime}
\author{N. Diaz-Naufal}
\affiliation{Dahlem Center for Complex Quantum Systems and Fachbereich Physik, Freie Universit\"{a}t Berlin, 14195 Berlin, Germany} 
\author{L. Deeg}
\affiliation{Institute for Quantum Optics and Quantum Information, Austrian Academy of Sciences, 6020 Innsbruck, Austria
}
\affiliation{Institute for Experimental Physics, University of Innsbruck, 6020 Innsbruck, Austria} 
\author{D. Zoepfl}
\affiliation{Institute for Quantum Optics and Quantum Information, Austrian Academy of Sciences, 6020 Innsbruck, Austria
}
\affiliation{Institute for Experimental Physics, University of Innsbruck, 6020 Innsbruck, Austria} 
\author{C. M. F. Schneider}
\affiliation{Institute for Quantum Optics and Quantum Information, Austrian Academy of Sciences, 6020 Innsbruck, Austria
}
\affiliation{Institute for Experimental Physics, University of Innsbruck, 6020 Innsbruck, Austria} 
\author{M. L. Juan}
\affiliation{Institut Quantique and D\'epartement de Physique,
Universit\'e de Sherbrooke, Sherbrooke, Qu\'ebec, J1K 2R1, Canada}
\author{G. Kirchmair}
\affiliation{Institute for Quantum Optics and Quantum Information, Austrian Academy of Sciences, 6020 Innsbruck, Austria
}
\affiliation{Institute for Experimental Physics, University of Innsbruck, 6020 Innsbruck, Austria} 
\author{A. Metelmann}
\affiliation{Dahlem Center for Complex Quantum Systems and Fachbereich Physik, Freie Universit\"{a}t Berlin, 14195 Berlin, Germany} 
\affiliation{Institute for Quantum Materials and Technology and Institute for Theory of Condensed Matter, Karlsruhe Institute of Technology,  
76131 Karlsruhe,
Germany}
\affiliation{Institut de Science et d’Ingénierie Supramoléculaires (ISIS, UMR7006), University of Strasbourg and CNRS}
 
\date{\today}
\begin{abstract} 
Dynamical backaction cooling has been demonstrated to be a successful method for achieving the motional quantum ground state of a mechanical oscillator in the resolved sideband regime, where the mechanical frequency is significantly larger than the cavity decay rate. Nevertheless, as mechanical systems increase in size, their frequencies naturally decrease, thus bringing them into the unresolved sideband regime, where the effectiveness of the sideband cooling approach decreases.
Here, we will demonstrate, however, that this cooling technique in the unresolved sideband regime can be significantly enhanced by utilizing a nonlinear  cavity as shown in the experimental work of Zoepfl \textit{et al.} \cite{Zoepfl2023}. The above arises due to the increased asymmetry between the cooling and heating processes, thereby improving the cooling efficiency.
\end{abstract}
\maketitle

\section{Introduction}
Besides emerging to create highly sensitive detectors for gravitational waves \cite{Barry_GWaves, Caves_GWaves}, the field of optomechanics has undergone enormous progress in recent years \cite{Aspelmeyer_Review}.   
Among the achievements are the detection of displacements below the standard quantum limit \cite{Teufel2009, Mason2019}, the preparation of nonclassical states of motion \cite{Wollman2015, Zivari2022}, and the demonstration of strong optomechanical coupling in both the microwave and optical regimes \cite{Teufel2011_StrongCoupling, delosRiosSommer2021_StrongCoupling}. 
Due to the controllable interaction between light and mechanics, optomechanics is considered a fundamental resource for realizing hybrid quantum devices of otherwise incompatible degrees of freedom for numerous applications in quantum technologies \cite{Barzanjeh2021}. In addition, optomechanical systems have been proven to be fundamental in shining a light on the quantum-to-classical transition \cite{Bai2017}.
However, a glimpse towards the underlying quantum behavior of macroscopic mechanical systems is hindered by the difficulty of cooling a mechanical mode to its quantum ground state \cite{Marquardt2008}.   
Even for systems at cryogenic temperatures, low-frequency mechanical systems require additional cooling due to their high thermal occupation. 

In the last decade, feedback \cite{Rossi2018, Tebbenjohanns_2020} and dynamical backaction cooling \cite{Teufel2011, Chan2011} protocols have been successful in achieving ground state cooling. The latter is based on the extraction of mechanical phonons by scattering incident drive photons to higher frequencies. This photon up-conversion (anti-Stokes) process competes with its counterpart (Stokes) process, which adds energy to the mechanical system. Dynamical backaction cooling works best in the \textit{resolved sideband} regime, where the mechanical resonant frequency exceeds the cavity decay rate $(\omega_m \gg \kappa)$. Within this regime, ground state cooling using a linear cavity has been achieved already over a decade ago \cite{Teufel2011, Chan2011}. However, increasing the size of mechanical systems leads to a natural decrease in their frequency, which brings them into the \textit{unresolved sideband} regime $(\omega_m \ll \kappa)$, where
Stokes and anti-Stokes processes become equally likely and hence, dynamical backaction cooling becomes ineffective. 
Nevertheless, \textit{measurement-based feedback} cooling is in this regime a more successful strategy \cite{Mancini_1998, Genes2008}, but it is limited by the imprecision of the readout \cite{Marquardt_Optomechanics}. 
Recently this limitation has been proposed to be overcome through the implementation of \textit{coherent feedback}, resulting in ground state cooling within the unresolved sideband regime \cite{Guo2022, Ernzer_2023}.   
Moreover, multiple alternative approaches have been discussed
such as using two cavity modes \cite{Yang2019, Liu2015} or frequency modulated light \cite{Wang2018}. Additionally, the engineering of entirely distinct coupling mechanisms has been proposed, such as a coupling of the mechanical system to the cavity decay rate \cite{Elste2009}, an   exciton polariton architecture \cite{Zambon2022}, 
or to additional two level systems \cite{Genes2009}. 

An alternative strategy to enhance cooling performance is the use of squeezed light, which can be generated either externally \cite{Asjad2016, Clark2017} or internally within the cavity \cite{Huang2009, Gan2019, Xiong2020, Lau2020}. The advantages and disadvantages of these approaches have been discussed in \cite{Asjad2019}.  
Such strategies are very promising but require a fine-tuning of the squeezing parameters to enable suppression of the Stokes scattering process and thus enhanced cooling. Moreover, the generation of squeezing requires a nonlinear element for the electromagnetic field, such as a $\chi_2$-medium in the THz-domain or Josephson-junction based circuit loops in the microwave domain.

Interestingly, a nonlinear cavity itself can be beneficial for back-action cooling of a mechanical mode in the unresolved sideband regime \cite{Nation, Laflamme2011}. 
In this paper we analyze in detail how the nonlinearity of a Kerr resonator in a magnetomechanical architecture can be beneficial for backaction cooling of a mechanical mode. With this, we are providing the theoretical background of the analysis used in our recent experimental work Zoepfl \textit{et al.} \cite{Zoepfl2023}. The latter work used a flux-mediated optomechanical coupling scheme \cite{Rodrigues2019, Zoepfl2020, Schmidt2020, Luschmann2022, Bothner2022}, where a mechanical oscillator is coupled magnetically to microwave photons in a superconducting LC circuit. Here, the displacement of the mechanical oscillator is converted into a magnetic flux that changes a superconducting quantum-interference device (SQUID), serving as flux-dependent inductance \cite{Nation, Shevchuk2017}. 

Crucially, besides being a magnetic field sensitive element, the SQUID is also a nonlinear element since its inductance depends on the number of photons circulating in the cavity, resulting in a typically unwanted Kerr nonlinearity within the optomechanical system. Despite limiting the driving power due to the emergence of a bistable regime, we show that a nonlinear cavity enables more efficient cooling than an identical linear system. Interestingly, the improvement in cooling performance arises particularly in the unresolved sideband regime.  

This paper is organized as follows: in Sec. \ref{sec_I} we introduce the system's Hamiltonian and study the resulting dynamics of the classical amplitude and quantum fluctuations. In Sec. \ref{sec_nonlinear_cav} we analyze the dynamical features of the nonlinear cavity in the absence of optomechanical interaction. Afterward, in Sec. \ref{sec_radiation_pressure_force}, we turn on the interaction between the cavity and mechanical modes and demonstrate how the cavity's nonlinearity can enhance the cooling efficiency. To study the influence of the nonlinear cavity on the cooling limits, in Sec. \ref{sec_cooling}, we derive the effective dynamics of the mechanical mode. Furthermore, in Sec. \ref{sec_towards_ground}, following the work of Asjad \textit{et al.} \cite{Asjad2016}, we demonstrate that by injecting squeezed vacuum the unwanted backaction heating can be suppressed. 
Here the advantages of using a nonlinear cavity becomes apparent as well, as less squeezing strength is required  to achieve the same level of backaction suppression if compared to an equivalent linear system.

\section{The model} \label{sec_I}
\subsection{Hamiltonian formulation} \label{sec_derivation_Hamiltonian}
\begin{figure}[t]
	\centering
	\includegraphics[width = 0.8\linewidth]{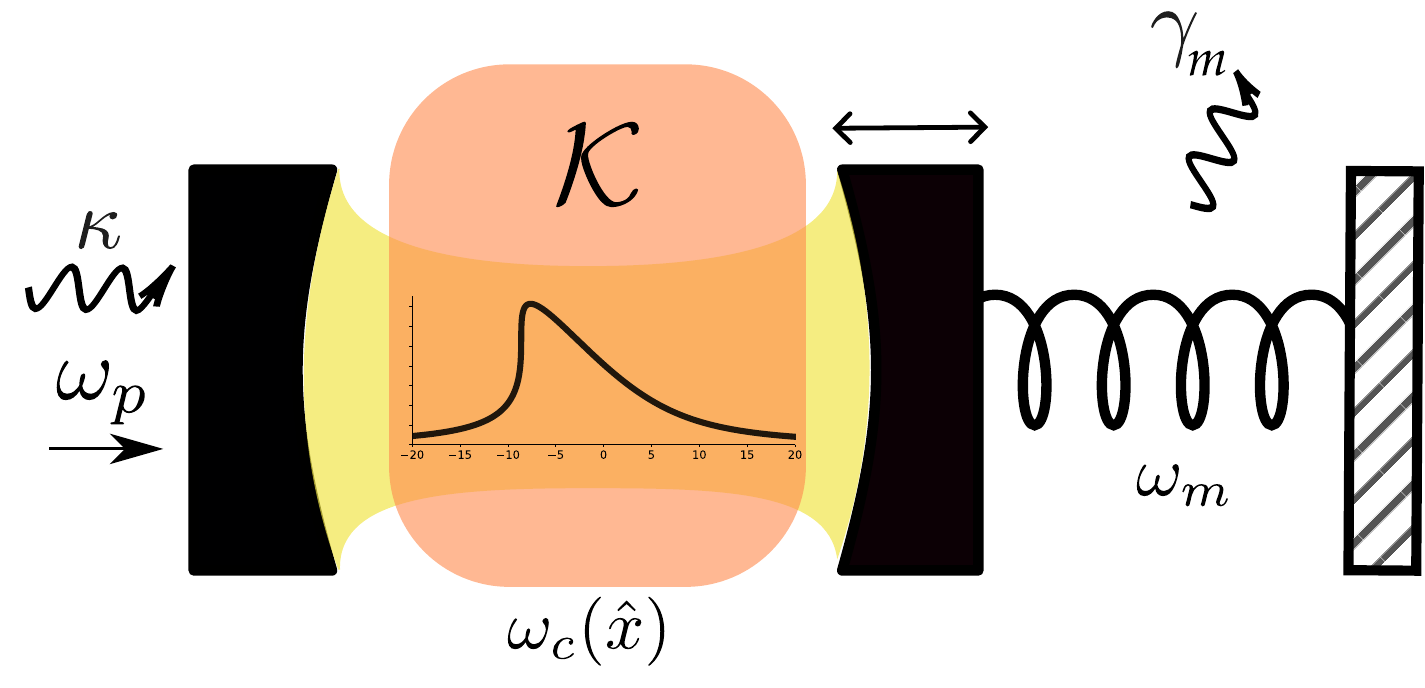}
	\caption{Optomechanical setup consisting of a driven Kerr-cavity coupled via radiation-pressure force to a mechanical resonator. Due to the presence of the nonlinearity, the average photon number circulating in the cavity exhibits a very prominent asymmetry. }
	\label{fig:model_sketch}
\end{figure}

The starting point of our discussion will be the Hamiltonian describing a mechanical oscillator parametrically coupled to a nonlinear cavity, where the latter corresponds to a Kerr-resonator which can be driven into a bistable regime as depicted in Fig.\ref{fig:model_sketch}. The Hamiltonian associated with the composite system is given as \cite{Nation}
\begin{align}
\hat{\mathcal{H}}_\text{tot} = \hat{H}_0 - \frac{\mathcal{K}}{12} \left( \hat{a} + \hat{a}^\dagger \right)^4 + \frac{g_0}{2} \left( \hat{a} + \hat{a}^\dagger \right)^2 \left( \hat{b} + \hat{b}^\dagger \right) + \hat{H}_\text{d},
\label{Hamiltonian_Nation}
\end{align}
where $\hat{a}(\hat{b})$ and $\hat{a}^\dagger(\hat{b}^\dagger)$ are annihilating and creating an excitation in the cavity (mechanical) mode, respectively. $\hat{H}_0 = \omega_c \hat{a}^\dagger \hat{a} + \omega_m \hat{b}^\dagger \hat{b}$ denotes the free Hamiltonian of the cavity and mechanical mode with corresponding resonance frequencies $\omega_c$ and $\omega_m$ (we set $\hbar=1$ throughout this work).  
$\mathcal{K}$ is the Kerr constant, typically assumed to be $\mathcal{K} > 0$, and $g_0 = g x_\text{zpf}$ denotes the bare optomechanical coupling strength composed by $x_\text{zpf}$ the zero-point-fluctuation amplitude of the mechanical resonator mode and $g = \partial\omega_c/ \partial x$ the cavity frequency shift per displacement. In addition, we include an external drive associated with the term $\hat{H}_\text{d} = \alpha_p e^{- i \omega_p t} \hat{a}^\dagger + \text{h.c}$, where $\omega_p$ and $\alpha_p$ are the drive frequency and classical amplitude, respectively.
Furthermore, assuming a weak enough coupling $g_0$ and nonlinearity $\mathcal{K}$, the system can be simplified by neglecting counter-rotating terms yielding the time-independent Hamiltonian
\begin{align}
\hat{\mathcal{H}} &= -\Delta\hat{a}^\dagger \hat{a} +  \omega_m \hat{b}^\dagger \hat{b} - \frac{\mathcal{K}}{2} \hat{a}^\dagger \hat{a}^\dagger \hat{a}\hat{a} + g_0 \hat{a}^\dagger \hat{a} \left( \hat{b} + \hat{b}^\dagger \right) + \hat{\mathcal{H}}_d,
\label{Hamiltonian_FirstApprox}
\end{align}
with the cavity mode rotated with respect to the drive frequency $\omega_p$, where $\Delta = \omega_p - \omega_c$ denotes the detuning and the Hamiltonian drive term reads $\hat{\mathcal{H}}_d = \alpha_p \hat{a}^\dagger + \text{h.c.}$.

To study both classical and quantum dynamics we introduce the so-called \textit{linearized} approximation, which is based on the assumption that the system is driven strongly at the drive's frequency $\omega_p$, such that its dynamics can be well regarded as only small fluctuations in the vicinity of the steady state mean values. Hence, we introduce the displacement operator $\hat{\mathcal{D}}(\eta) = \exp{(\eta \hat{o}^\dagger - \eta^* \hat{o})}$ with $\eta = \langle \hat{o} \rangle$ as an average amplitude and $\hat{o}$ the fluctuations around it. This allows us to transform the Hamiltonian Eq.\ \eqref{Hamiltonian_FirstApprox} to a displaced frame of the cavity and mechanical mode. The above results in the effective description of the coherent dynamics of the fluctuations given by the Hamiltonian (see App. \ref{sec_derivation_Hamiltonian})
\begin{align} 
	\begin{split}
		\hat{\mathcal{H}}_\text{eff} &= -\tilde{\Delta} \hat{d}^\dagger \hat{d} + \omega_m \hat{b}^\dagger \hat{b} - \frac{1}{2} \left[ \Lambda \hat{d}^\dagger \hat{d}^\dagger + \Lambda^* \hat{d} \hat{d} \right] 
		\\
		&+ \left( G \hat{d}^\dagger + G^*\hat{d} \right) \left( \hat{b} + \hat{b}^\dagger \right)	,
	\end{split}
	\label{Hamiltonian_Rotated}
\end{align}
with the effective detuning $\tilde{\Delta} = \Delta + 2 |\Lambda|$. This time-independent Hamiltonian is characterized by the single-mode squeezing strength $\Lambda  = \mathcal{K} \bar{n}_c e^{2i \phi_c}$, and the photon enhanced optomechanical coupling strength $G = g_0 \sqrt{\bar{n}_c} \,e^{i \phi_c}$ with $\phi_c \in \left[0, 2\pi\right)$. Here, the average coherent amplitude of the cavity field is given by $\alpha = \sqrt{\bar{n}_c} e^{i \phi_c}$ with its phase $\phi_c$  and the intracavity photon number $\bar{n}_c$.    

The first two terms in $\hat{\mathcal{H}}_\text{eff}$ describe the modified free Hamiltonian, where the presence of the Kerr nonlinearity introduces a photon-dependent frequency shift. The third term describes a parametric amplification process induced by the Kerr nonlinearity, which plays a crucial role in squeezing generation \cite{Carmichael_1984_DPA}. The final term describes the usual linearized optomechanical interaction, which
combines the process of swapping excitation with the process of two-mode squeezing between both modes. In the resolved sideband regime it is possible to select either of these processes depending on the respective frequency of the drive tone. However, in the unresolved regime, both processes are contributing and in our analysis, we account for both of them.

\subsection{Classical dynamics}	 \label{sec_classical_dynamics}
To achieve a sufficiently large cooling rate, we are interested in driving strengths that lead to a large average number of intracavity photons $\bar{n}_c$, but at the same time do not result in multistable solutions in the classical cavity dynamics \cite{Laflamme2011}. 
Hence, our first focus is on understanding the classical dynamics of the system.
We start from the nonlinear Hamiltonian Eq.\ \eqref{Hamiltonian_FirstApprox} and assume that the cavity is coupled to an external waveguide with rate $\kappa$. We use standard \textit{input-output} theory \cite{input_output} to obtain   
\begin{table}
\centering
\begin{tabular}{ ||p{0.45\linewidth} p{0.45\linewidth}|| }
 \hline
 Parameter & Value \\
 \hline
 Mechanical frequency & $\omega_m/2\pi = 0.3 \text{ MHz}$    \\
 Mechanical linewidth & $\gamma_m/2\pi = 0.5 \text{ Hz}$  \\
 Cavity linewidth & $\kappa/2\pi = 3\text{ MHz}$ \\
 Kerr strength & $\mathcal{K}/2\pi = 0.16 \text{ MHz}$ \\
 Optomechanical coupling & $g_0/2\pi = 1.7 \text{ kHz}$\\ 
 Input power & $\bar{n}_\text{in,crit} \equiv  0.9999999 \bar{n}_\text{in,bi}$\\ 
 [1ex]
 \hline
\end{tabular}
\caption{Default values chosen for the parameters in the figures throughout the paper, if not otherwise indicated. These values are similar to those from our recent experimental work \cite{Zoepfl2023}. 
}
\label{table_param}
\end{table}
the equation of motion 
for the average coherent cavity amplitude $\alpha$, from whose steady state solution we can deduce the average cavity photon number (see App. \ref{sec_explicit_bistable})  
\begin{align}
	\bar{n}_c \left[ \left( \Delta + \mathcal{K}_\text{eff} \bar{n}_c \right)^2 + \left( \frac{\kappa}{2}\right)^2 \right] = \kappa \bar{n}_\text{in},
	\label{cubic_photon}
\end{align}
with the effective Kerr constant
\begin{align}
	\mathcal{K}_\text{eff} \equiv  \mathcal{K} + \frac{2 g_0^2 \omega_m }{\omega^2_m + \frac{\gamma_m^2}{4}},
	\label{effective_kerr}
\end{align}  
and the input photon flux $\bar{n}_\text{in}$ (in units of photons per second). The cavity and mechanical decay rates are labelled as $\kappa$ and $\gamma_m$, respectively. Note, the effective Kerr Eq.\ \eqref{effective_kerr} comprises both, the intrinsic cavity and the optomechanically-induced nonlinearities \cite{Nunnenkamp_Kerr_Medium, Dorsel_Bistability}, respectively. The latter will henceforth be referred to as the \textit{mechanical Kerr}.   

The average number of photons within the cavity described by the cubic equation Eq.\ \eqref{cubic_photon} provides insight into the cavity's behaviour. For low drive strengths, a single real solution exists. However, for stronger drive power a bifurcation occurs and a regime of bistability arises, where two stable states appear within a certain range of detunings.
\begin{figure}[t]
	\centering 
 \includegraphics[width = \linewidth]{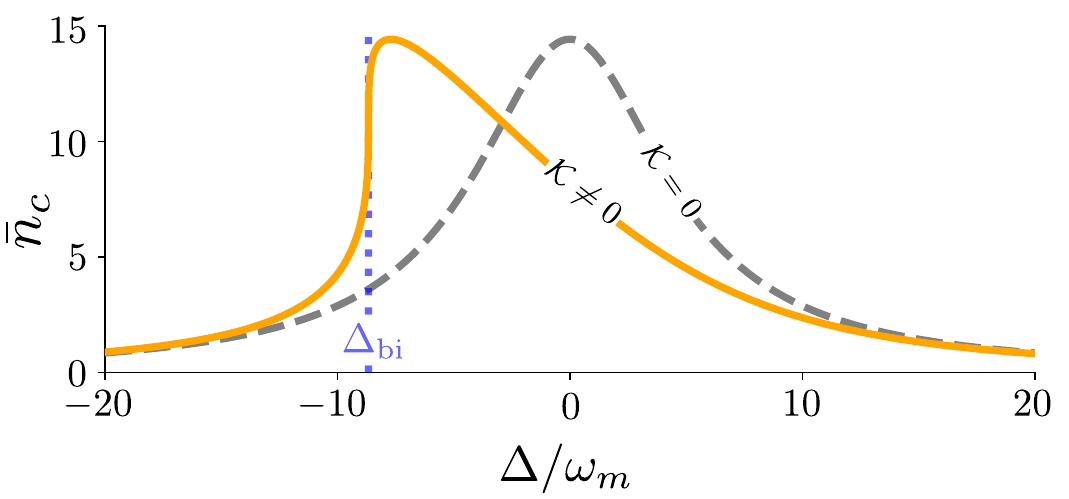}
	\caption{Average cavity photon number as a function of the detuning $\Delta = \omega_p - \omega_c$. The orange solid line shows the intracavity photon number obtained using a nonlinear cavity ($\mathcal{K} \neq 0$) near the point of bifurcation, namely at a drive amplitude of $\bar{n}_\text{in} = \bar{n}_\text{in,crit}$. In contrast, the grey-dashed line results from the linear cavity setup ($\mathcal{K} = 0$) at the same input power. The vertical dotted line denotes the critical detuning $\Delta_\text{bi}$, where the system becomes bistable. 
 }
	\label{fig:Intra_Photon}
\end{figure}
The bifurcation takes place at a point in parameter space 
where the first derivative of $\bar{n}_c$ with respect to $\Delta$ diverges, which happens at a single point $\Delta = \Delta_\text{bi}$. Furthermore, by imposing the requirement of continuity on the transition between the two regions, we arrive at
\begin{align}
	\begin{split}
		\Delta_\text{bi} = -\frac{\sqrt{3}\kappa}{2}, \quad
		\bar{n}_\text{bi} =	\frac{\kappa}{\sqrt{3} \mathcal{K}_\text{eff}},
	\end{split}
	\label{bifurcation_universal}
\end{align}
which correspond to the universal values at bifurcation. Moreover, upon substituting the aforementioned relations into Eq.~\eqref{cubic_photon}, the critical drive amplitude $\bar{n}_\text{in,bi}$ at which bifurcation occurs can be obtained, i.e.
\begin{align}
	\bar{n}_\text{in,bi} = \frac{\kappa^2}{3\sqrt{3} \mathcal{K}_\text{eff}}.
	\label{critical_drive}
\end{align}

\begin{figure}[t]
	\centering 
    \includegraphics[width = \linewidth]{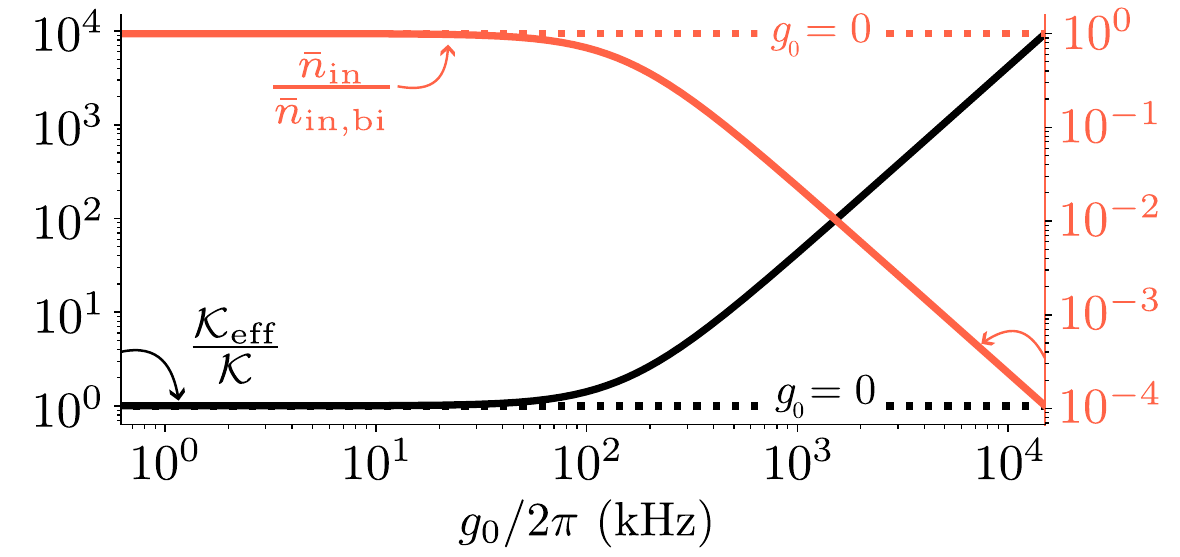}
	\caption{The black solid line depicts the effective Kerr nonlinearity defined in Eq.~\eqref{effective_kerr} normalized by $\mathcal{K}$ as a function of the optomechanical coupling strength $g_0/2\pi$. For given parameters (see Table
 \ref{table_param}), the intrinsic cavity nonlinearity dominates for weak coupling strengths, and the critical input power (red line) is dominated by $\mathcal{K}$.  Increasing the coupling strength further the mechanical Kerr kicks in and the critical input power decreases as the effective Kerr is enhanced. Dashed lines (color) correspond to the case without the induced mechanical Kerr, i.e. $g_0 = 0$.}
	\label{fig:critical_Input_eff_Kerr}
\end{figure}

Consequently, for drive amplitudes slightly below the critical driving threshold, given by Eq.~\eqref{critical_drive}, the average photon number  
exhibits a single-valued solution with respect to the cavity detuning, with a significant gradient at $\Delta_\text{bi}$, as represented by Eq.~\eqref{bifurcation_universal} and illustrated in Fig. \ref{fig:Intra_Photon}. However, for the chosen parameters (see Table \ref{table_param}), an identical linear cavity ($\mathcal{K} = 0$) driven at the same input power would show the conventional Lorentzian distribution for the average photon number, despite the presence of the mechanical Kerr.
Hence, for weak coupling strengths, the intrinsic nonlinearity of the cavity dominates over the mechanical Kerr nonlinearity, as depicted in Fig.~\ref{fig:critical_Input_eff_Kerr}. 

\subsection{Dynamics of the fluctuations}
To fully understand the implications of the interaction between the mechanical and the cavity mode, and thus the limit for cooling of the mechanical resonator, we analyze next the dynamics of the fluctuations. We start from the effective Hamiltonian Eq.~\eqref{Hamiltonian_Rotated} to derive the dissipative dynamics of the fluctuations described by the quantum Langevin equation
\begin{align}
	\frac{d}{dt}\vec{A} = \mathbf{M} \vec{A} 
 - \mathbf{K} \vec{A}_\text{in},
	\label{Quantum_Langevin}
\end{align}
where we defined the vector containing all mode operators $\vec{A} = [ \hat{d}, \hat{d}^\dagger, \hat{b}, \hat{b}^\dagger ]^T$ and the dynamical matrix $\mathbf{M}$ as
\begin{align}
	\mathbf{M} = 
	\begin{pmatrix}
		i \tilde{\Delta} - \frac{\kappa}{2} & i \Lambda & -iG & -iG \\
		-i\Lambda^* & -i \tilde{\Delta} - \frac{\kappa}{2} & iG^* & iG^* \\
		-iG^* & -iG & -i\omega_m - \frac{\gamma_m}{2} & 0 \\
		iG^* & iG & 0 & i\omega_m - \frac{\gamma_m}{2}
	\end{pmatrix}.
\end{align}
In Eq.~\eqref{Quantum_Langevin}, the vector $\vec{A}_\text{in} = [  \hat{d}_\text{in},  \hat{d}^\dagger_\text{in}, \hat{b}_\text{in},  \hat{b}_\text{in}^\dagger]^T$ represents the cavity and mechanical input noises, whereas the decay rates are encoded in the diagonal matrix $\mathbf{K} = \text{diag}(\sqrt{\kappa}, \sqrt{\kappa}, \sqrt{\gamma_m}, \sqrt{\gamma_m})$. The vacuum noise operator $\hat{d}_\text{in}$ satisfies $\langle \hat{d}_\text{in}(t)\hat{d}^\dagger_\text{in}(t') \rangle = \delta(t - t')$ and $\langle \hat{d}^\dagger_\text{in}(t)\hat{d}_\text{in}(t') \rangle = 0$. Analogously, the noise operator $\hat{b}_\text{in}$ describes coupling to a Markovian reservoir at temperature $T$, as represented by the following correlators $\langle \hat{b}_\text{in}(t)\hat{b}^\dagger_\text{in}(t') \rangle = (\bar{n}_m^T + 1)\delta(t - t')$ and
$\langle \hat{b}^\dagger_\text{in}(t)\hat{b}_\text{in}(t') \rangle = \bar{n}_m^T \delta(t - t')$. Furthermore, in the absence of any other coupling, the bath leads to the emergence of a thermal state, characterized by a mean occupation number $\bar{n}_m^T = [\exp\{\omega_m/k_B T\} -1 ]^{-1}$ for the mechanical oscillator.

In the following sections, we investigate the consequences resulting from the optomechanical interaction. This can be accomplished by solving the linearized equations of motion in Eq.~\eqref{Quantum_Langevin} in the frequency domain, from which we can deduce the relevant noise spectra and subsequently the mechanical occupation. As the response of the nonlinear cavity will determine the cavity cooling rate we first examine the cavity in the absence of coupling to the mechanical mode.

\section{Nonlinear cavity} \label{sec_nonlinear_cav}
\begin{figure}[t]
	\centering 
 \includegraphics[width = \linewidth]{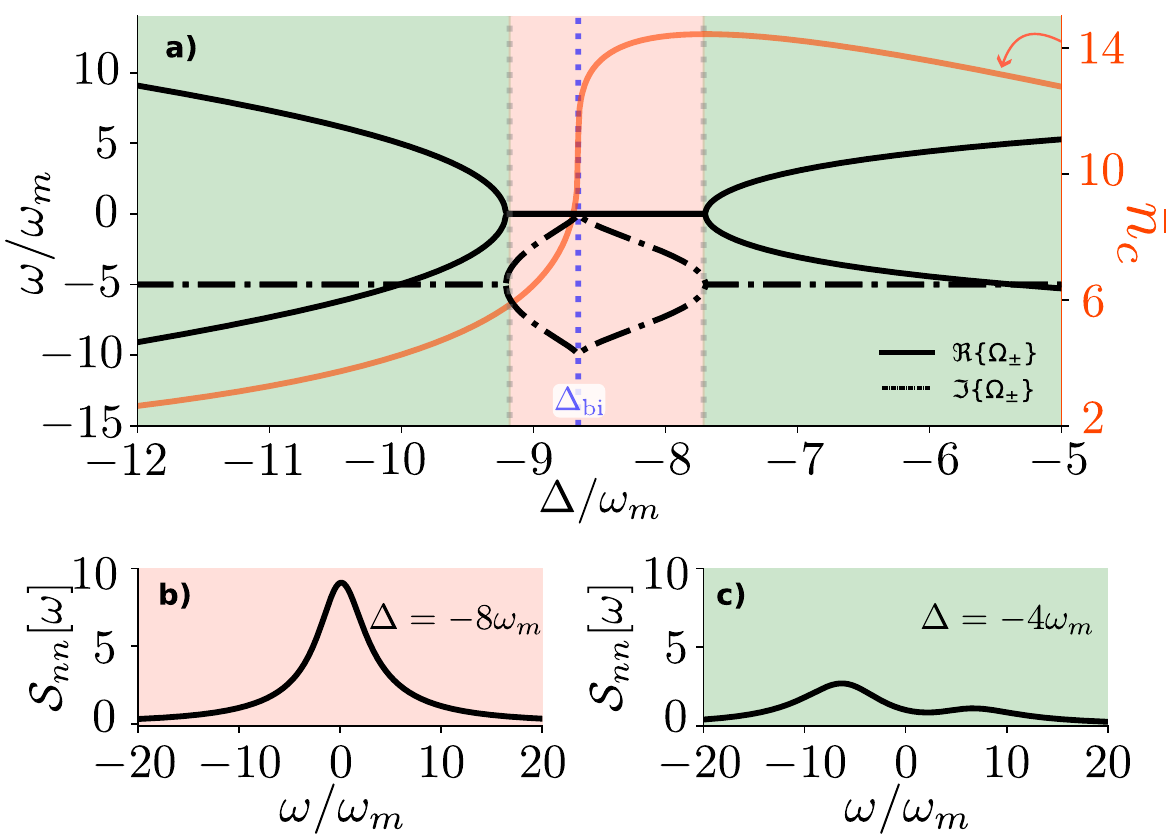}
	\caption{\textbf{a)} Poles of the driven Kerr cavity in the absence of coupling to the mechanical mode as a function of the detuning. Here, we depict the real (solid) and imaginary (dashed-dotted) part of the poles given in Eq.\ \eqref{poles}, which are associated with the system's resonance frequency and decay, respectively. Exceptional points are shown by the vertical grey-dotted lines, which delimit the interval where the system features two distinct decay rates at a single resonant frequency (red shaded and \textbf{b}). The EPs are found at the intersection of $\Lambda = \mathcal{K} \bar{n}_c$ with the lines $-\Delta/3$ and $-\Delta$. Outside this interval, the system exhibits split resonance frequencies for a single decay rate (green shaded) leading to a double peak structure in the photon number spectrum as shown in \textbf{c)}.  Note that, the decay rates' maxima/minima occur precisely at the bifurcation detuning $\Delta_\text{bi}$ shown here by the vertical blue line.}
	\label{fig:Intra_Poles}
\end{figure}

From Eq.~\eqref{Quantum_Langevin} we find that in the absence of optomechanical interaction, the dynamics of the Kerr cavity fluctuations are described by the Hamiltonian of a detuned parametric amplifier \cite{Carmichael_1984_DPA} yielding  
\begin{align}
	\hat{d}[\omega] = - \sqrt{\kappa} \, \tilde{\mathcal{X}}_c[\omega] \left( \hat{d}_\text{in}[\omega] + i \Lambda \mathcal{X}^{*}_c[-\omega] \hat{d}^\dagger_\text{in} [\omega] \right),
	\label{kerr_cavity_fluctuations}
\end{align}
where we defined the driven Kerr cavity susceptibility as
\begin{align}
	\widetilde{\mathcal{X}}_c[\omega] = \frac{\mathcal{X}_{c}[\omega]}{1 - |\Lambda|^2\,  \mathcal{X}_c[\omega] \mathcal{X}^*_c[-\omega]},
	\label{suscept_cavity}
\end{align} 
with $\mathcal{X}^{-1}_c[\omega] = -i(\omega + \tilde{\Delta}) + \kappa/2$. The added idler noise,
i.e., the second term in Eq.~\eqref{kerr_cavity_fluctuations},
and the modification of the cavity susceptibility  
will result in 
an asymmetrical shape of the photon number spectrum $\mathcal{S}_{nn}[\omega] = \int_{-\infty}^\infty dt \,e^{i\omega t} \langle (\hat{a}^\dagger \hat{a})(t) (\hat{a}^\dagger \hat{a})(0) \rangle$ given explicitly as (see App. \ref{app_force_spectra}) 
\begin{align}
	\mathcal{S}_{nn}[\omega] = \frac{\bar{n}_c \kappa \left( \left[-\tilde{\Delta} + \omega  + |\Lambda| \right]^2 + \frac{\kappa^2}{4}\right)}{\left[\tilde{\Delta}^2 - \omega^2 + \frac{\kappa^2}{4} - |\Lambda|^2  \right]^2 + \kappa^2 \omega^2}.
	\label{photon_spectra}
\end{align} 
The asymmetric response of the nonlinear cavity is an important characteristic for the system's ability to absorb energy. This becomes clearer when we analyze the spectrum in more detail. First
we consider the poles of the spectrum, 
which can be extracted from the denominator of Eq.~\eqref{photon_spectra}: 
\begin{align}
	\Omega_{c,\pm} = -i \frac{\kappa}{2} \pm \sqrt{\left( \Delta + 3 |\Lambda| \right)\left( \Delta + |\Lambda| \right)},
	\label{poles}
\end{align}
here, the real (imaginary) part of the poles represents the resonance frequencies (damping rates) of the system. In Fig.~\ref{fig:Intra_Poles}a) the poles are plotted as a function of the bare detuning $\Delta$, and we can identify two distinct regions. For $\Delta < -3 |\Lambda|$ and $\Delta > - |\Lambda|$    
(green shaded area) the dynamics of the nonlinear cavity are governed by two resonant frequencies $\Re \{ \tilde{\Omega}_{\pm}\}$ and a single decay rate resulting in the formation of two peaks in the photon number spectrum as shown in Fig.~\ref{fig:Intra_Poles}c). On the other hand,  for
 $ -3 |\Lambda| < \Delta < - |\Lambda|$
(red shaded area) the situation is reversed and the eigenvalues show degenerate imaginary parts and split dissipation rates $\Im \{ \tilde{\Omega}_{\pm} \}$ as in Fig.~\ref{fig:Intra_Poles}a). These two regions are delimited by exceptional points (EPs) (dotted vertical lines), which are common degeneracies in open quantum systems \cite{Mller2008}. At an EP the system eigenvalues become degenerate and both eigenvectors coalesce due to the vanishing square root in Eq.~\eqref{poles}. The EPs occur for the condition 
$\Delta_{\pm} = -(2 \pm 1) |\Lambda|$, however, the exact detuning points  are not straightforwardly determined as $|\Lambda|$ is a function of the average photon number in Eq.~\eqref{cubic_photon} which itself depends on the detuning.

We expect that the dynamical backaction onto the mechanical mode is proportional to the variation of the photon number \cite{Nation, Laflamme2011}, hence the steepness of the slope in the average photon number, cf. orange line in Fig.~\ref{fig:Intra_Poles}a), 
suggests that the relevant detuning regime for optimal cooling is the red region in Fig.~\ref{fig:Intra_Poles}a). 
In this region, and close to the critical detuning $\Delta_\text{bi}$, we observe the interesting effect that one imaginary part of the pole approaches zero, which can be interpreted as a critical slowing down of the cavity. The latter translates into a narrowing of the photon number spectrum, reminiscent of the gain-bandwidth limitation of a parametric amplifier \cite{Clerk_Noise}. We deduce the exact condition for the extrema values via
\begin{align}
		\frac{d}{d\bar{n}_c} \Im\left\{\tilde{\Omega}_{\pm} \right\} = 0 
 \; \;  \Rightarrow \; \;
  \frac{\Bar{n}_c}{\Delta} = - \frac{2}{3 \mathcal{K}},
\end{align} 
which is equivalent to $\bar{n}_\text{bi}/\Delta_\text{bi}$ given in Eq.~\eqref{bifurcation_universal} for $\mathcal{K}_\text{eff} \approx\mathcal{K}$. This indicates that, as a consequence of the cavity nonlinearity, the cavity features a minimum/maximum decay rate exactly at the critical detuning in Eq.~\eqref{bifurcation_universal}, which arises due to the pronounced variation of the photon number with respect to the detuning as shown in Fig. \ref{fig:Intra_Poles}a). 

However, in the detuning range enclosed by the EP's we solely have a single resonance peak at $\omega =0$ (in this rotated frame).  
Crucially, changing the detuning within this range will not move the photon number spectrum along the frequency axis, which seems to be an issue at first glance. For this we have to remember the linear case, the optimal detuning in the resolved sideband regime is obtained for $\Delta = -\omega_m$ where the peak of the spectrum is located at the red-sideband $\omega = \omega_m$. For a symmetric spectrum not being able to move the spectrum in frequency space would be an issue, as no net damping could emerge. However, here is where the nonlinear response of the cavity comes into play, i.e., the asymmetry of the spectrum is essential to obtain cooling.

\begin{figure}[t]
	\centering 
 \includegraphics[width = \linewidth]{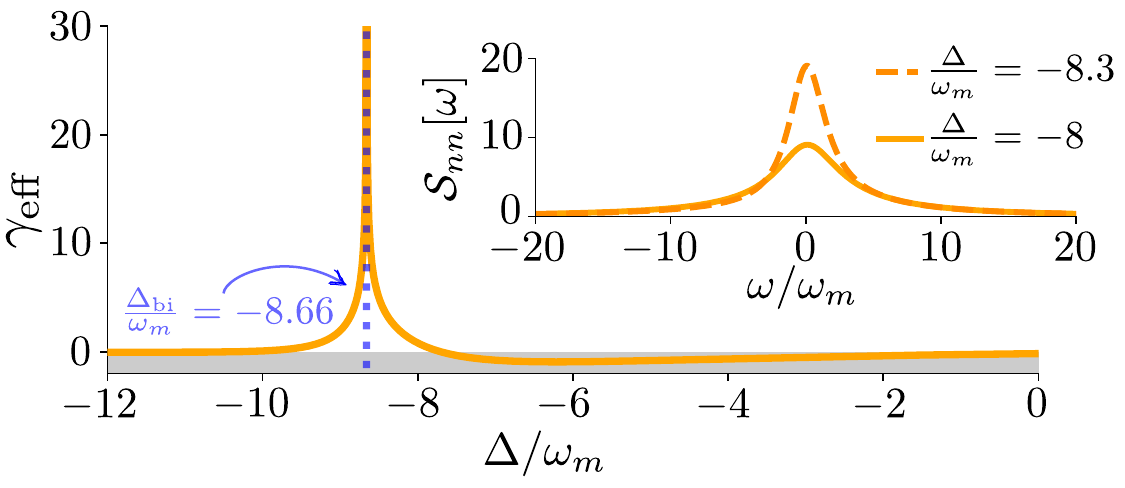}
	\caption{Effective skewness of the photon number spectrum of a nonlinear cavity as a function of the detuning. Inset shows the corresponding photon number spectrum as a function of the driving frequency for detunings approaching the critical value $\Delta_\text{bi}$ (vertical dashed-blue line). These spectra follow a slightly asymmetric Lorentzian distribution, whose maximum asymmetry is found when the cavity is driven at $\Delta = \Delta_\text{bi}$. Thus, as the cavity is driven close to the critical detuning the photon number spectrum becomes increasingly peaked and asymmetric, so for illustrative purposes we chose $\Delta = - 8 \omega_m$ (solid) and $\Delta = - 8.3 \omega_m$ (dashed).}
	\label{fig:skewness}
\end{figure}

 
To quantify the asymmetry we use the \textit{skewness}, a measure which in general is utilized to characterize the asymmetry of a probability distribution. Fig.~\ref{fig:skewness} depicts the photon number spectrum's \textit{effective skewness} $\gamma_\text{eff}$
\footnote{Here, we numerically obtained the \textit{moment-based skewness} using the following function
\begin{align}
\gamma_1 = \frac{\sum_{i=1}^{n} (x_i - \mu)^3}{n \sigma^3},
\end{align}
where $x$ is an array of data points with length $n$ from the distribution, and $\mu$ and $\sigma$ are the mean and standard deviation of the distribution, respectively. Furthermore, since the Lorentz distribution is heavy-tailed and therefore does not have finite moments, i.e. well-defined mean and variance, here we calculated the \textit{truncated moment-based skewness}. This involves the data set to a finite interval, calculating the moments of the truncated distribution, and then using these moments to determine the skewness of the truncated distribution. Hence, for a certain interval of detunings $\Delta$, we calculated the skewness of the photon number spectrum given in Eq.\ \eqref{photon_spectra} over a truncated set of driving frequencies $\omega$. That means, for a $\Delta/\omega_m \in [-12, 0]$ the array $x$ is given by the set of elements fulfilling $x = \left\{ \mathcal{S}_{nn}[\omega] \mid \omega/\omega_m \in [-100,100] \right\}$.   In addition, in Fig.~\ref{fig:skewness} we show the \textit{effective skewness}, which results from subtracting the constant skewness obtained using a linear cavity to the associated one of a nonlinear cavity, that is we do $\gamma_\text{eff}(\Delta) = \gamma_1^{\mathcal{K} \neq 0}(\Delta) - \gamma_1^{\mathcal{K} = 0}$ with $\gamma_1^{\mathcal{K} = 0} = 3.48$.}
as a function of the cavity detuning. The maximum asymmetry occurs exactly at the critical detuning $\Delta_\text{bi}$, a detuning, 
where we also expect the largest fluctuations due to the steepness of the slope in the average photon number, cf. Fig.~\ref{fig:Intra_Photon}. 
A positive skewness corresponds to a slightly stronger decline of the photon number spectrum for negative frequencies and an extended tail for positive frequencies. In other words, for $\gamma_\text{eff} > 0$ we have 
$\mathcal{S}_{nn}[\Omega] > \mathcal{S}_{nn}[-\Omega]$
with $\Omega> 0$, which is a crucial feature for cooling the mechanics later on, as the positive (negative) frequency part of the spectrum $\mathcal{S}_{nn}[(-)\Omega] $ can be interpreted as the ability of the system to absorb (emit) energy \cite{Clerk_Noise}. For cooling we want the cavity to absorb energy from the mechanical mode and hence a positive skewness indicates the detuning regime of cooling. Moreover, we can already identify the required condition for the nonlinear cavity to absorb energy as
\begin{align}\label{EqCondCool}
    \mathcal{S}_{nn}[\Omega] -  \mathcal{S}_{nn}[-\Omega]
    > 0 
    \hspace{0.2cm}
    \text{for}
     \hspace{0.2cm}
       \Delta    < - |\Lambda| ,
\end{align} 
thus we expect that cooling will be possible for $\Delta  < - |\Lambda|$, which is distinct from the linear case which requires a negative detuning, i.e. cooling for $\Delta<0$ if $\mathcal K =0$.

In summary, the photon number spectrum $\mathcal{S}_{nn}[\omega]$ differs clearly from the typical Lorentzian form of a linear cavity \cite{Aspelmeyer_Review}. While a Lorentzian is fully symmetric around its maximum, the spectrum for the nonlinear cavity is asymmetric. We see that this asymmetric aspect is fundamental for the ability to cool the mechanical system.  
Next, we will demonstrate how the nonlinearity of a cavity can enhance the cooling efficiency of a mechanical resonator in the unresolved sideband regime by exploiting its asymmetrical response.

\section{Dynamical backaction} \label{sec_radiation_pressure_force} 
In the unresolved sideband regime, where the cavity decay rate far exceeds the mechanical frequency $(\kappa \gg \omega_m)$, the cavity field adapts \textit{almost} instantaneously to the oscillator's position. However, when we consider the cavity's finite response time, a phase lag arises between the radiation pressure force and the mechanical motion. This retardation effect causes the force to become out of sync with the oscillator's displacement, resulting in a net energy extraction from the mechanical mode \cite{Marquardt2008}.   
Due to the presence of the intrinsic Kerr nonlinearity, this effect is more pronounced in the interval of split decay rates (red shaded) in Fig.~\ref{fig:Intra_Poles}. A a consequence of this nonlinearity the cavity ring-down time 
has its maximum exactly at the point of bifurcation $\Delta_\text{bi}$ as shown in Sec.~\ref{sec_nonlinear_cav}. Hence, approaching the critical detuning leads to an effective reduction of cavity linewidth, bringing the cavity into the resolved regime. However, cooling results solely from the imbalance between the Stokes and anti-Stokes processes, which is enhanced by the presence of the nonlinearity as depicted in Fig.~\ref{fig:skewness}.

In the weak coupling regime, $g_0 \ll \kappa$, the quantum theory of optomechanical cooling can be studied within a perturbative picture \cite{Marquardt2008}, which is best explained in the Raman-scattering framework. Here, photons from below the cavity resonance frequency are scattered upwards in frequency to enter the cavity resonance, absorbing in the process a phonon from the mechanical mode. These anti-Stokes processes occur at a rate $\Gamma_\text{AS}$. On the other hand, the Stokes processes, where photons return red-shifted through the creation of one excitation in the mechanics, happen at a rate $\Gamma_\text{S}$. Therefore, cooling can be viewed to originate from the imbalance between the Stokes and anti-Stokes scattering, such that the full optomechanical damping is the net downward rate $\Gamma_\text{opt} = \Gamma_\text{AS} - \Gamma_\text{S}$.

The rates $\Gamma_\text{AS,S}$ can be obtained using Fermi's Golden Rule applied to the photon-pressure interaction between the mechanical oscillator and the photonic cavity, namely $\hat{H}_\text{int} = \hat{F} \hat{x}$ with the radiation pressure force $\hat{F} = g_0 \hat{a}^\dagger \hat{a}$ \cite{Clerk_Noise}. Hence, within the weak coupling limit, the Stokes and anti-Stokes rates are given by 
\begin{align}
	\Gamma_\text{S,AS} = \mathcal{S}_{FF}[\mp \omega_m] = g_0^2 \mathcal{S}_{nn}[\mp \omega_m] ,
	\label{stokes_anti_stokes}
\end{align}
with the photon number spectrum given in Eq.\ \eqref{photon_spectra}, the optical damping becomes
\begin{align}
    \Gamma_{\text{opt}} =  
  \frac{ 4   
  g_{0}^2\bar{n}_c   
  \left( |\Lambda|
       - \tilde{\Delta} \right) 
    \kappa  \omega_m   
  }{\left[\tilde{\Delta}^2 - \omega_m^2 + \frac{\kappa^2}{4} - |\Lambda|^2  \right]^2 + \kappa^2 \omega_m^2},
\end{align}
from which we recover the cooling condition in Eq.~\eqref{EqCondCool}, i.e., we have positive optical damping for
$ |\Lambda|   - \tilde{\Delta} 
  = - (\Delta + |\Lambda|)>  0 $. 
Moreover, the expression for the optical damping indicates that a \textit{dynamical backaction evasion} scheme can be realized when $\tilde{\Delta} = |\Lambda|$, which corresponds exactly to the condition of one of the EPs discussed in Sec.\ \ref{sec_nonlinear_cav}. At this point the rates for heating and cooling are equal and the optical damping vanishes and the mechanics only suffers from backaction due to the fluctuations of the cavity scaling with $\Gamma_{S}$.
This suggests that a nonlinear cavity located at the EP could be used for measurement of the oscillator's motion at the standard quantum limit \cite{Marquardt_Optomechanics, Teufel2009}. 

\begin{figure}[t]
	\centering
    \includegraphics[width = \linewidth]{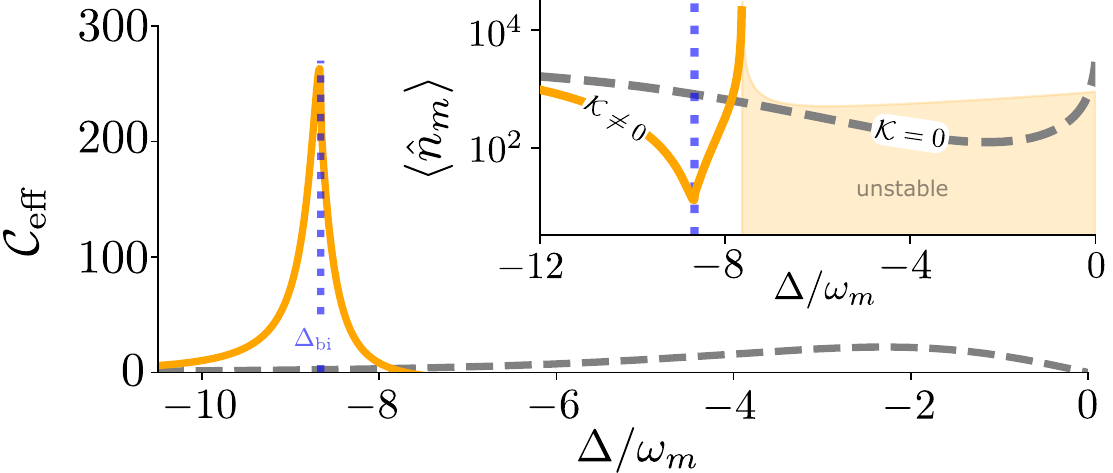}
	\caption{Effective cooperativity as a function of the optical detuning for a nonlinear (orange line) and linear (dashed grey line) cavity.
 The asymmetric shape of the photon spectrum (see Fig.~\ref{fig:skewness}) due to the Kerr nonlinearity results
 in the imbalance between the Stokes and anti-Stokes rates. This results in an increase of the cavity damping, primarily around the critical detuning. Inset shows the average phonon number as a function of the cavity detuning, where the minimum occupation occurs at the point of maximum induced damping. For the parameters given in Table \ref{table_param}, we find that the cooling capabilities of a nonlinear system outperform an equivalent but linear system by more than an order of magnitude.}
	\label{fig:cooperativity}
\end{figure}

However, we a here interested in cooling of the mechanical mode, and thus we are aiming for maximal dynamical backaction. As discussed in Sec.\ref{sec_nonlinear_cav}, the asymmetry of the photon number spectrum induced by the presence of the Kerr nonlinearity produces an appreciable imbalance between the Stokes and anti-Stokes rates and consequently to an increase in the \textit{effective cooperativity} $ \mathcal{C}_\text{eff} \equiv \Gamma_\text{opt}/\gamma_m$. This can be seen in Fig.~\ref{fig:cooperativity} where the maximum cavity damping coincides with the point of maximum asymmetry, at $\Delta_\text{bi}$, where the cavity ring-down time is maximized. At this bifurcation point, the circulating photon number within the nonlinear cavity exhibits a significant variation, thereby enhancing the cooling process \cite{Zoepfl2023}.

It has been proposed that intra-cavity squeezing can be utilized to remove the backaction onto the mechanical mode \cite{Huang2009, Gan2019, Xiong2020, Lau2020}, i.e., by achieving 
$\Gamma_{S} = 0$. This procedure requires a careful matching  of the optomechanical interaction and squeezing strengths in their magnitude, and most importantly, as well as in their phase.  
Here we have a single phase that originates from the cavity amplitude ($\phi_c$) and which enters our coherent dynamics in Eq.~\eqref{Hamiltonian_Rotated}. This phase can be gauged away and hence we have no phase-independence of our optomechanical and squeezing processes.  
This prevents us from using it for the suppression of the unwanted backaction heating. However, as we will study later, injecting squeezed vacuum light into the cavity can be employed to suppress the Stokes rate given in Eq.\ \eqref{stokes_anti_stokes}, thereby allowing the system to surpass the backaction limit in its cooling capabilities \cite{Vitali_16}.

\section{Cooling} \label{sec_cooling}
In order to investigate the limits of cooling in our setup, we analyze the influence of the nonlinear cavity on the dynamics of the mechanical mode by deriving the effective average mechanical occupation. For this we solve the equations of motion \eqref{Quantum_Langevin} for the $b$-mode to find an effective dynamical description of the mechanical oscillator. 
Defining the vector $\vec{\matr{B}}[\omega] = [\hat{b}[\omega], \hat{b}^\dagger[\omega]]^T$ we obtain
\begin{align}
	\boldsymbol{\chi}^{-1}_\text{m,eff}[\omega] 
	\vec{\matr{B}}[\omega]
	= - \sqrt{\gamma_m} \,
	\vec{\matr{B}}_{\text{in,eff}}[\omega],
\label{Eff_Quantum_Langevin_Mechanics}
\end{align} 
with $\vec{\matr{B}}_{\text{in,eff}}[\omega] = [\hat{B}_\text{in}[\omega], \hat{B}_\text{in}^\dagger[\omega]]^T$
containing the modified noise impinging on the mechanical mode:
\begin{align}
	\begin{split}
			\hat{B}_\text{in}[\omega] &=  
   \hat{b}_\text{in}[\omega] + \frac{i}{\sqrt{\gamma_m}}
   \left[ 
   \eta[\omega] \hat{d}^\dagger_\text{in}[\omega] + \eta^*[-\omega] \hat{d}_\text{in}[\omega]
   \right] 
		\end{split}
\end{align} 
with the coefficient
\begin{align}
    \eta[\omega] = -\frac{\sqrt{\kappa}}{\mathcal{N}[\omega]} \left( G\mathcal{X}_{c}^{-1}[\omega] + i G^* \Lambda \right),
\end{align}
where we introduced $\mathcal{N}[\omega] = \mathcal{X}_c^{*-1}[-\omega] \mathcal{X}_c^{-1}[\omega] - |\Lambda|^2$.  
Additionally, the response of the mechanical mode is encoded in the susceptibility matrix 
\begin{align}
	\begin{split}
		\boldsymbol{\chi}^{-1}_\text{m,eff}[\omega]= 
		\begin{pmatrix}
			\mathcal{X}^{-1}_\text{m}[\omega] - i \Sigma_c[\omega] &
			- i \Sigma_c[\omega]
			\\
			i \Sigma_c[\omega] & \mathcal{X}^{*-1}_\text{m}[-\omega] + i \Sigma_c[\omega]
		\end{pmatrix}
	\end{split},
\label{susceptibilit_effective}
\end{align}
with the cavity self energy
\begin{align}
	\Sigma_c[\omega] \equiv \frac{2 |G|^2\left( |\Lambda| -\tilde{\Delta} \right)}{( -i \omega + \frac{\kappa}{2})^2 + \tilde{\Delta}^2 - |\Lambda|^2 },
	\label{cavity_self_energy}
\end{align} 
and the mechanical susceptibility $\mathcal{X}^{-1}_m[\omega] = -i(\omega - \omega_m) + \gamma_m/2$.
Due to the optomechanical interaction, the effective susceptibility in Eq.\ \eqref{susceptibilit_effective} contains the modified mechanical frequency and damping rate $\omega_m -  \Re \left\{ \Sigma_c[\omega] \right\}$ and $\gamma_m + 2 \Im \left\{ \Sigma_c[\omega] \right\}$, respectively. 
In addition to the frequency shift ($\Re \left\{ \Sigma_c[\omega] \right\}$) and the damping ($\Im \left\{ \Sigma_c[\omega] \right\}$), the cavity also induces single-mode squeezing in the mechanical oscillator with effective coupling strength $ \Sigma_c[\omega]$.
As pointed out before, a cancellation of dynamical backaction can be achieved when the nonlinear cavity is located at the EP $\tilde{\Delta} = |\Lambda|$, meaning that the cavity self-energy $\Sigma_c[\omega]$  vanishes. In contrast, if a linear cavity is employed, an equivalent effect arises on resonance $\Delta = 0$. 
 
For sufficient weak coupling $(\kappa \gg g_0)$, the cavity self energy can be evaluated at the mechanical resonance frequency, namely $\Sigma_c[\omega] \approx \Sigma_c[\omega_m]$. Hence, assuming a high-Q mechanical oscillator the mechanical noise spectrum takes the rather simple form 
\begin{align}
	\begin{split}
		\mathcal{S}_{bb}[\omega] &=  \frac{\gamma_m \left|\mathcal{X}_m^{*-1}[-\omega] + i \Sigma_c[\omega_m]\right|^2 }{\left|\left(\omega - \Omega_{m,+}\right)\left(\omega - \Omega_{m,-}\right)\right|^2}  \bar{n}_m^T 
		\\
		&+ \frac{\gamma_m\left|\Sigma_c[\omega_m]\right|^2 }{\left|\left(\omega - \Omega_{m,+}\right)\left(\omega - \Omega_{m,-}\right)\right|^2}  \left(\bar{n}_m^T +1 \right) 
		\\
		&+ 
		\frac{\left|\mathcal{X}_m^{*-1}[-\omega] \right|^2}{\left|\left(\omega - \Omega_{m,+}\right)\left(\omega - \Omega_{m,-}\right)\right|^2} \, \Gamma_\text{S}, 
	\end{split}
	\label{spectra_quality_factor}
\end{align}
with the poles $\Omega_{m,\pm} = - i \gamma_m/2 \pm \sqrt{\omega_m^2 - 2\omega_m \Sigma_c[\omega_m]}$ and the Stokes rate given by Eq.\ \eqref{stokes_anti_stokes}.  
The first two terms in the final expression represent the modified mechanical noise contributions, which are associated with both the thermal and vacuum fluctuations of the oscillator's mode.
Note that the contribution proportional to $\bar{n}_m^T + 1$ arises exclusively due to the optomechanical interaction yielding the squeezing of the mechanics. 
Lastly, the final term represents the \textit{cavity backaction noise}, which originates solely from the interaction with the cavity and contributes an additional source of noise.

The average occupation of the mechanical mode can be obtained via the integration of the mechanical noise spectrum
\begin{align}
	\bar{n}_m = \int \frac{d\omega}{2 \pi} \mathcal{S}_{bb}[\omega],
	\label{mech_occupation_spectra}
\end{align}
with the noise spectrum given in Eq.\ \eqref{spectra_quality_factor}. Since for the current parameters we have $\Sigma_c[\omega_m]/\omega_m \ll 1$ 
the integral can be solved analytically.  
The above implies that the off-diagonal elements of the effective mechanical susceptibility given in Eq.\ \eqref{susceptibilit_effective} can be neglected and with this, the induced squeezing, described by  non-resonant processes. 

Using Eq.\ \eqref{mech_occupation_spectra} and assuming a high-Q mechanical oscillator we can approximate the average mechanical occupation as
\begin{align}
	\langle \hat{n}_m \rangle \approx  \frac{\bar{n}_m^T}{  \mathcal{C}_\text{eff} + 1} + \frac{\mathcal{C}_\text{eff}}{  \mathcal{C}_\text{eff} + 1} \frac{ \left(\omega_m - \Delta_\text{eff}\right)^2 + \frac{\kappa^2}{4} }{4 \omega_m \Delta_\text{eff}},
	\label{mech_occ_steady}
\end{align}
with  
the detuning $\Delta_\text{eff} \equiv |\Lambda| - \tilde{\Delta}$. The first term in Eq.\ \eqref{mech_occ_steady}
corresponds to a modified thermal occupation, whereas the second term comes from the unwanted cavity backaction heating. Note that in the absence of optomechanical interaction, the mechanical occupation is on average its thermal occupation, i.e. $\bar{n}_m^T$.

An equivalent description of the cooling, as indicated in Sec. \ref{sec_radiation_pressure_force}, is provided by an effective master equation for the mechanical mode, which results in the steady-state average mechanical occupation  (see Sec. \ref{sec_equivalence})
\begin{align}
	\langle \hat{n}_m \rangle_\text{s} = \frac{\gamma_m \bar{n}_m^T}{\gamma_m + \Gamma_\text{opt}} + \frac{\Gamma_\text{S}}{\gamma_m + \Gamma_\text{opt}}
	\label{mech_occ_steady_state}
\end{align}
with the Stokes and anti-Stokes rates given by Eq.\ \eqref{stokes_anti_stokes}. Thus, the substitution of Eq.\ \eqref{stokes_anti_stokes} into Eq.\ \eqref{mech_occ_steady_state} coincides with the approximate result obtained in Eq.\ \eqref{mech_occ_steady}.

\begin{figure}[t]
	\centering
    \includegraphics[width = \linewidth]{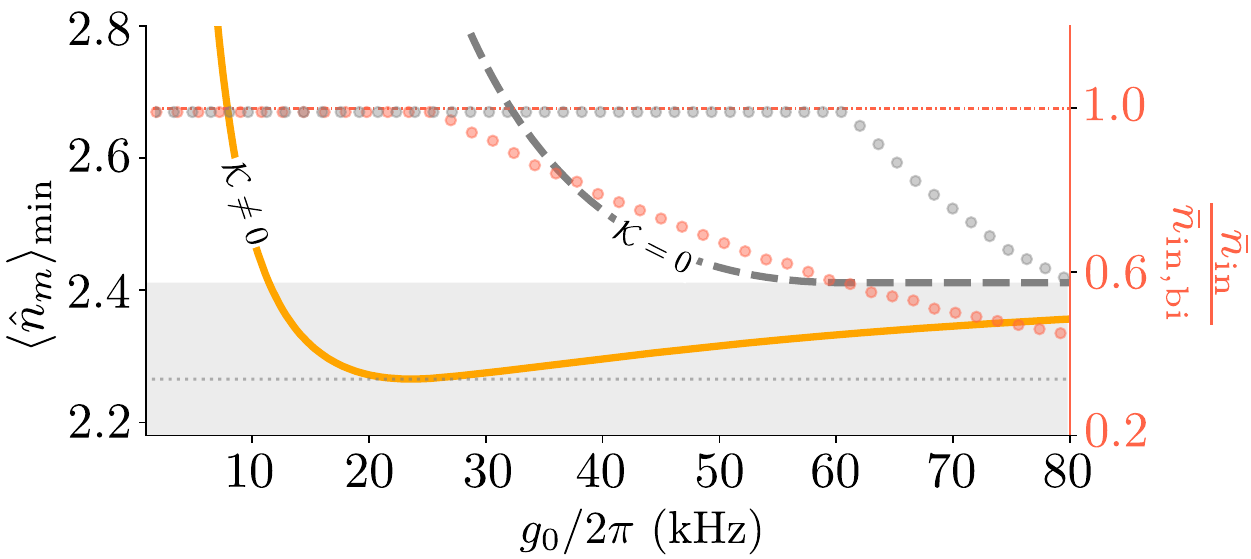}
	\caption{Lowest phonon number reached as a function of the optomechanical coupling strength. The use of a nonlinear cavity (orange line) outperforms an identical but linear setup (grey dashed line) even for weaker coupling strengths. For each value of coupling strength we have used the optimal input power below bifurcation (red/gray dotted line for $\mathcal K \neq 0/ \mathcal K =0$). Note, the optimal input power drops down for larger coupling strengths minimizing the backaction contribution. 
    The grey-shaded region shows the minimal phonon occupation reached for large $g_0$ in the linear case. Remaining parameters as given in Table \ref{table_param}.}
	\label{fig:Minima_Coupling}
\end{figure}
From Eq.~\eqref{mech_occ_steady} we find that for high effective cooperativity, the minimal  phonon number is set by the cavity backaction yielding
\begin{align}
	\bar{n}_\text{BA} \equiv \lim_{\mathcal{C}_\text{eff} \to \infty} \bar{n}_m = \frac{\left(\omega_m - \Delta_\text{eff}\right)^2 + \frac{\kappa^2}{4}}{4 \omega_m \Delta_\text{eff}},
	\label{cavity_backaction}
\end{align}
which by replacing $\Delta_\text{eff}  \rightarrow -\Delta$ coincides with the expression of the lowest achievable phonon number 
if a linear cavity is employed \cite{Aspelmeyer_Review}. 
Hence, the ultimate cooling limit is imposed by the cavity backaction just as in the linear case. However, we can still determine the optimal detuning in Eq.\ \eqref{cavity_backaction} as
$\Delta_\text{eff} \to \sqrt{\kappa^2/4 + \omega_m^2}$, which minimizes $\bar{n}_\text{BA}$ yielding $\bar{n}_\text{BA}^\text{min} = (\sqrt{\kappa^2/\omega_m^2 + 4} - 2)/4$. Thus, ground state cooling, i.e. $\bar{n}_\text{BA}^\text{min} < 1$, is only possible when  $\omega_m/\kappa > (4\sqrt{2})^{-1} \approx 0.177$, which coincides with the case of a linear cavity.

The experimental work by Zoepfl \textit{et al.} \cite{Zoepfl2023} worked with similar parameters as those provided in Table \ref{table_param}, and with $\omega_m/\kappa = 0.1 $ in the unresolved sideband regime. A high effective cooperativity  enables in particular the suppression of the thermal occupation as we see from the first term of Eq.\ \eqref{mech_occ_steady}. Operating below the bifurcation point of the nonlinear cavity, the driving strength is constrained, and with the given parameters, we can reach an effective cooperativity of $\mathcal{C}_\text{eff} \approx 264$, as shown in Fig.\ \ref{fig:cooperativity}. In contrast, for equivalent parameters and driving strength, a linear cavity achieves an effective cooperativity of only $\mathcal{C}_\text{eff} \approx 22$. This demonstrates that, under the given conditions, the use of a nonlinear cavity leads to an improvement of an order of magnitude in cooling performance. To reach a comparable cooling limit, a linear cavity would require nearly 12 times more input power. This is feasible because the linear system is not limited by the drive power in the same way as a nonlinear cavity. Nevertheless, this highlights the superior cooling efficiency of a nonlinear cavity compared to a linear cavity at low input powers.

With the parameters given in Table\ \ref{table_param}, the mechanical occupation can be suppressed from 2778 thermal phonons down to $\bar{n}_m = 12.66$, which is about $220$ times lower than the thermal occupation. This result agrees with the cooling performance demonstrated experimentally in \cite{Zoepfl2023}. According to Eq.\ \eqref{mech_occ_steady}, this remaining occupation is constituted of the modified thermal and  backaction contributions, with the latter accounting for $2.74$ phonons. In contrast, with the same parameters and identical input power, a linear cavity achieves a cooling limit of $123.33$ phonons. This demonstrates that, under the given conditions, a nonlinear cavity improves cooling performance by nearly an order of magnitude as depicted in the inset of Fig.\ \ref{fig:cooperativity}. 
Additionally, it is also worth noting the possibility to operate above bifurcation as in the work of L. Deeg \textit{et al.} \cite{Deeg_2024}.


We now loosen up the constraints imposed by the parameters  given in Table \ref{table_param}, and 
explore the lowest phonon occupation reached with the nonlinear setup as we increase the optomechanical coupling strength.  
The minimum phonon number, depicted in Fig.~\ref{fig:Minima_Coupling}, is achieved by optimizing the input power for each optomechanical coupling strength, with the constraint that the system remains in the monostable regime. Moreover, for larger coupling strengths the thermal contribution can be suppressed more efficiently, and optimal input power then means to work with a lower power value to reduce the backaction.  
The nonlinear setup allows cooling down to $\bar{n}_m = 2.26$ for a coupling strength of $g_0/2\pi = 23.28 \text{ kHz}$, which 
represents a $26\%$ improvement over the linear setup for identical parameters yielding $\bar{n}_m = 3.06$ (not shown in the graph).  
Within the chosen range of input powers the linear minimum can be further suppressed down to
2.41 phonons by increasing the coupling strength significantly.
Thus, cooling with a linear cavity requires a larger optomechanical coupling strength or a substantial increase in the input power. 
In contrast, the nonlinear setup provides a more efficient cooling at lower $g_0$ even exceeding the linear minimum of 2.41 phonons.
However, for the parameters given in Table\ \ref{table_param}, the minimum backaction, $\bar{n}_\text{BA}^\text{min} \approx 2.049$, is only possible with a linear cavity, since it has no bifurcation constraints. 
Nevertheless, reaching this limits requires for e.g. $g_0/2\pi = 15\text{ kHz}$ a substantial increase in power strength of $\bar{n}_\text{in} \approx 10^2 \bar{n}_\text{in,bi}$. 
 
\begin{figure}[t]
	\centering 
    \includegraphics[width = \linewidth]{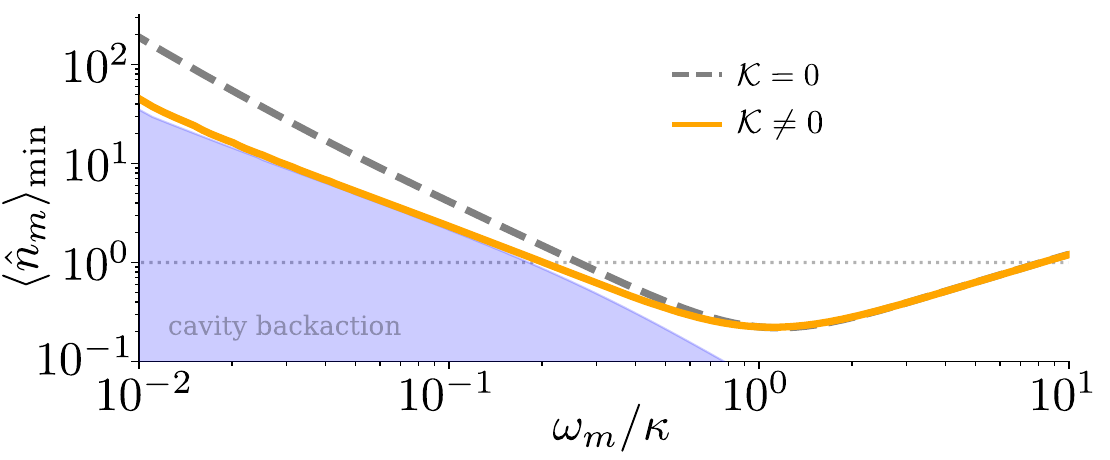}
	\caption{The lowest phonon occupation is shown as a function of the resolved sideband parameter, $\omega_m/\kappa$. In the unresolved sideband regime, the nonlinear cavity (orange curve) achieves better cooling performance compared to a linear cavity (dashed grey curve) at the same input power. For the given parameters, with an optomechanical coupling strength of $g_0/2\pi = 15\text{ kHz}$, the mechanical occupation is constrained by the cavity's backaction (blue shaded area).}
	\label{fig:Minima_Resolved}
\end{figure}

To further study the benefits of the nonlinear setup we analyse the lowest mechanical occupation achieved as a function of the resolved parameter $\omega_m/\kappa$. In Figure ~\ref{fig:Minima_Resolved} we compare the cooling performance of the nonlinear  with the conventional linear system for fixed $g_0/2\pi = 15\text{ kHz}$ and both equally driven at $\bar{n}_\text{in} = \bar{n}_\text{in,crit}$. For this optomechanical coupling strength, the mechanical occupation reached using the nonlinear cavity lies close to the minimum phonon number provided by the cavity backaction (blue shaded area). In contrast, the linear system does not reach the backaction limit for the given input power. Remarkably, we find that the more unresolved our system is, the better our improvement will be using a nonlinear cavity.  
However, as we approach the resolved sideband regime the advantage of the nonlinear setup diminishes showing similar results as the linear system.
As expected from our discussion of the backaction limit below Eq.\ \eqref{cavity_backaction}, for a resolved sideband parameter of around
$ \omega_m/\kappa \approx 0.2$ an occupation below one is achieved.  
A linear system requires here a similar value, but notably, the ratio $g_{0}/\kappa$ has to be roughly twice as high for the same input power, see App. \ref{app_minOcc} for further details. 
Note that for $\omega_m/\kappa > 1$, we observe an increase in temperature because the driving power for the linear case is not optimized. In contrast, for the nonlinear setup, the temperature increase is constrained by the critical input power.
  
\begin{figure}[t]
	\centering  
    \includegraphics[width = \linewidth]{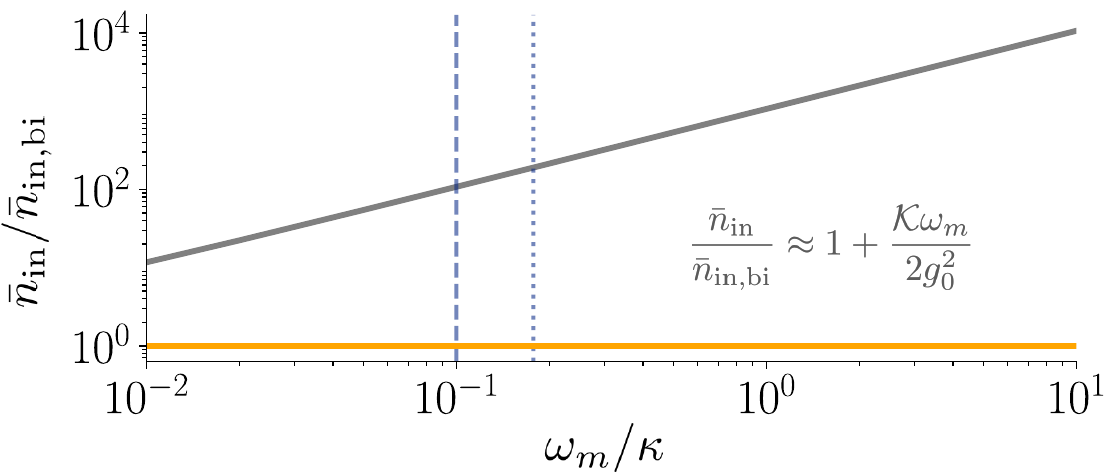}
	\caption{Required input power for optimal cooling as a function of the resolved sideband parameter, $\omega_m/\kappa$. A linear cavity (grey line) operating in the monostable requires substantially more input power to reach the backaction limit compared to a nonlinear system (orange line). The dashed vertical line represents experimental parameters, while the dotted vertical line marks the resolved sideband parameter from which ground state cooling becomes feasible. For the given parameters, with an optomechanical coupling strength of $g_0/2\pi = 15\text{ kHz}$.}
	\label{fig:Optimal_Power}
\end{figure}

So far we have mainly focused on a comparison with an equally strong driven linear cavity, highlighting the improved performance of a nonlinear setup in the unresolved sideband regime. However, by driving stronger a linear cavity can in general reach the backaction limit. In Figure \ref{fig:Optimal_Power} we show the required input power for such an optimal cooling. It becomes clearly apparent that substantially more input power is required for the linear setup. Restricting the linear cavity to the monostable regime, i.e., to powers below the bistable regime induced by the mechanical Kerr, we find that the optimal input power is at $ \bar{n}_\text{in} = 0.99 \bar{n}_\text{in,bi}^{\mathcal K = 0}$ and with Eq. \eqref{critical_drive} we obtain
\begin{align}
 \frac{ \bar{n}_\text{in} }
   {\bar{n}_\text{in,bi}}
   \approx 
    \frac{\bar{n}_\text{in,bi}^{\mathcal K = 0}} {\bar{n}_\text{in,bi}}
   = 1 +
  \frac{ \mathcal K } {   \mathcal K_m}
   \approx 
   1 + \frac{\mathcal K \omega_m}{2 g_{0}^2},
\end{align}
where we used the definition of the mechanical Kerr in Eq. \eqref{effective_kerr} in the last step. Thus, the input power exhibits a linear dependence with the mechanical frequency. 
 
Figure~\ref{fig:Minima_Resolved} also makes clear that deep in the unresolved sideband regime there is no way around the cavity backaction. Here alternative protocols are required, utilizing for example feedback or squeezing.
In the next section, we will provide a pathway towards the ground state utilizing the squeezing tactic for our nonlinear setup, following the linear protocol developed by Asjad \textit{et al.} \cite{Vitali_16}.

\section{Towards the ground state} \label{sec_towards_ground}
In Sec.~\ref{sec_cooling} we demonstrated that in the unresolved sideband regime a nonlinear cavity outperforms a linear system in its efficiency. Nevertheless, despite this improvement, cooling to the ground state is still constrained by the cavity backaction as shown in Fig.~\ref{fig:Minima_Resolved}. 
To overcome this limitation and cool below the backaction limit the use of optomechanical induced transparency \cite{Ojanen2014} and pulsed cooling schemes \cite{Liu2013} have for example been proposed. 
Alternatively, the use of squeezing, generated inside or outside the cavity, has been studied to improve the cooling performance in an optomechanical system \cite{Vitali_16, Clark2017, Vitali_19, Clark2016, Monsel_2021}. 
In this section we discuss how such a squeezing approach is straightforwardly transferable to the nonlinear system. Following \cite{Vitali_16}, we show that driving a nonlinear optomechanical cavity with squeezed vacuum allows us to suppress the Stokes process and thereby strongly reduce the unwanted cavity backaction.  
\begin{figure}[t]
	\centering 
    \includegraphics[width = \linewidth]{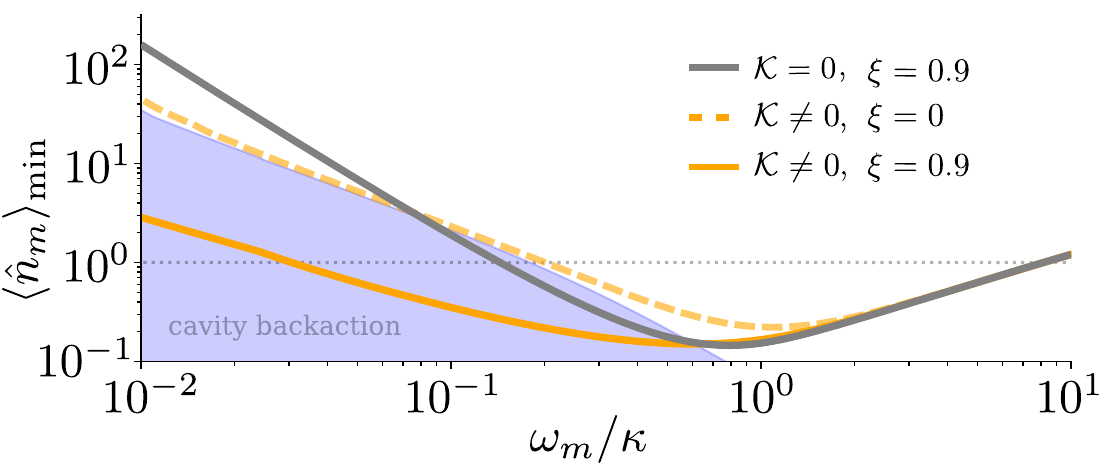}
	\caption{Lowest phonon occupation as a function of the resolved sideband parameter $\omega_m/\kappa$ at $g_0/2\pi = 15\text{ kHz}$. A nonlinear cavity driven with only vacuum noise (dashed-orange line) is limited by the unwanted cavity backaction, which can be suppressed when squeezed vacuum enters the cavity (solid-orange line). This suppression is as well observed for the linear case
    for the same parameters (solid-grey line), but is less successful  in comparison to the nonlinear case. An advantage of the nonlinear system which decreases as $\omega_m$ increases.
 }
\label{fig:Minima_Resolved_squeezing}
\end{figure}

Formally, pure squeezed states arise through the action of the \textit{squeezing operator} $\hat{S}(r) = \exp\left[ r e^{i \theta} \hat{a}^2/2 + \text{h.c} \right]$ onto the vacuum state of a bosonic mode $\hat a$, here $r \in [0,\infty)$ and $\theta \in [0, 2\pi]$ denote the squeezing strength and angle, respectively. Experimentally, squeezed states can be generated by injecting vacuum into a degenerate parametric amplifier (DPA) \cite{Carmichael_1984_DPA, Andersen2016}.   
Assuming that the output of the DPA is cascaded into the optomechanical system yields the new cavity noise operators 
\begin{align}
	\begin{split}
		\langle \hat{d}^\dagger_\text{in}(t) \hat{d}_\text{in}(t')\rangle
		&= \xi N_s \delta(t- t'),
		\\
		\langle \hat{d}_\text{in}(t) \hat{d}_\text{in}(t')\rangle
		&= \xi M_s e^{- 2i \varphi}\delta(t- t'),
	\end{split}
	\label{White_noise}
\end{align}
with
\begin{align}
	N_s &= \frac{\left(\kappa_+^2 - \kappa_-^2\right)^2}{4 \kappa_+^2 \kappa_-^2}, \quad 
	M_s = \frac{\kappa_+^4 - \kappa_-^4}{4 \kappa_+^2 \kappa_-^2},
	\label{White_noise_correl}
\end{align}
with $\kappa_\pm = \kappa/2 \pm |\chi|$, where we utilize $\chi = |\chi|e^{-2i \varphi}$ as the squeezing parameter and $\varphi$ as the squeezing angle
(see App. \ref{app_squeezing} for details). Furthermore, we introduce
a scaling parameter $\xi$ which accounts for intrinsic losses in the system, this parameter quantifies how effectively the squeezed vacuum is produced and routed towards the nonlinear cavity. 
We refer to this parameter as the purity of the squeezing as $\xi = 1$ denotes pure squeezing and $\xi = 0$ vacuum noise only. 
Note, that for the latter case the noise correlators in Eq.\ \eqref{White_noise} vanish as expected for vacuum.
Using these correlators  we can derive the radiation pressure force spectrum $\mathcal{S}_{FF}[\omega]$  (the explicit expression can be found in App. \ref{app_squeezing}). 
From the force spectrum we can extract the total damping rate $\Gamma_\text{tot} = \mathcal{S}_{FF}[\omega_m] - \mathcal{S}_{FF}[-\omega_m]$ and determine the cavity backaction limit $\bar{n}_\text{BA} = \Gamma_\text{S}/\Gamma_\text{tot}$. To note is, that the total damping rate $\Gamma_\text{tot}$ is independent of the input squeezed light and can therefore not be used to improve the cooling performance. Nevertheless, we can still minimize the Stokes process and consequently reduce the cavity backaction by choosing the optimal squeezing phase obtained from $d/d\varphi \;\Gamma_\text{S} =0 $. 
Using this optimal squeezing phase the cavity backaction  
becomes
\begin{align}
	\bar{n}_\text{BA}^\text{s} = \bar{n}_\text{BA}\left\{ 1 + \left[ 1 + \frac{1}{\wp^2}\right] \xi N_s
	- \frac{2 \xi M_s}{\wp}  \right\},
	\label{Backlimit_Main}
\end{align}
with $\bar{n}_\text{BA}$ given by Eq.\ \eqref{cavity_backaction} and 
\begin{align}
	\wp^2 = \frac{ \left[\Delta_\text{eff} - \omega_m  \right]^2 + \frac{\kappa^2}{4}}{ \left[\Delta_\text{eff} + \omega_m   \right]^2 + \frac{\kappa^2}{4}}.
    \label{wp_value}
\end{align}
We can minimize the backaction in Eq.\ \eqref{Backlimit_Main} over 
the squeezing parameter incorporated in $N_s$ and find that for $\wp = \sqrt{N_s/(N_s + 1)}$ we obtain
\begin{align}
	\bar{n}_\text{BA}^\text{s} =    \bar{n}_\text{BA}\left( 1 -  \xi  \right).
 \label{squeezed_BA}
\end{align}
As expected, in the absence of squeezing we recover Eq.\ \eqref{cavity_backaction}, but for $\xi = 1$  we can fully suppress the unwanted cavity backaction.  
In Fig.~\ref{fig:Minima_Resolved_squeezing} we show the lowest phonon number achieved 
when squeezed vacuum is injected into the cavity. As explained before, the injected squeezing results in the possibility of suppressing the Stokes scattering process, thus reducing the cavity backaction heating. In this way, the use of a squeezed input  allows us to exceed the cooling limits obtained when only vacuum noise enters the nonlinear cavity. This improvement is also appearing when squeezed vacuum  is injected into a linear cavity as depicted by the grey line in Fig.~\ref{fig:Minima_Resolved_squeezing}. However, as in Sec.~\ref{sec_cooling}, in the unresolved sideband regime a nonlinear cavity outperforms an equivalent linear system when both are equally driven. Furthermore, for a nonlinear cavity at $g_0/2\pi = 15\text{ kHz}$ and with $\xi = 0.9$, ground state cooling is achieved for $\omega_m/\kappa = 0.03$, whereas for a linear cavity it requires $\omega_m/\kappa = 0.144$, which demonstrated again the benefits of a nonlinear cavity in the unresolved sideband regime.  
 
\begin{figure}[t]
	\centering 
    \includegraphics[width = \linewidth]{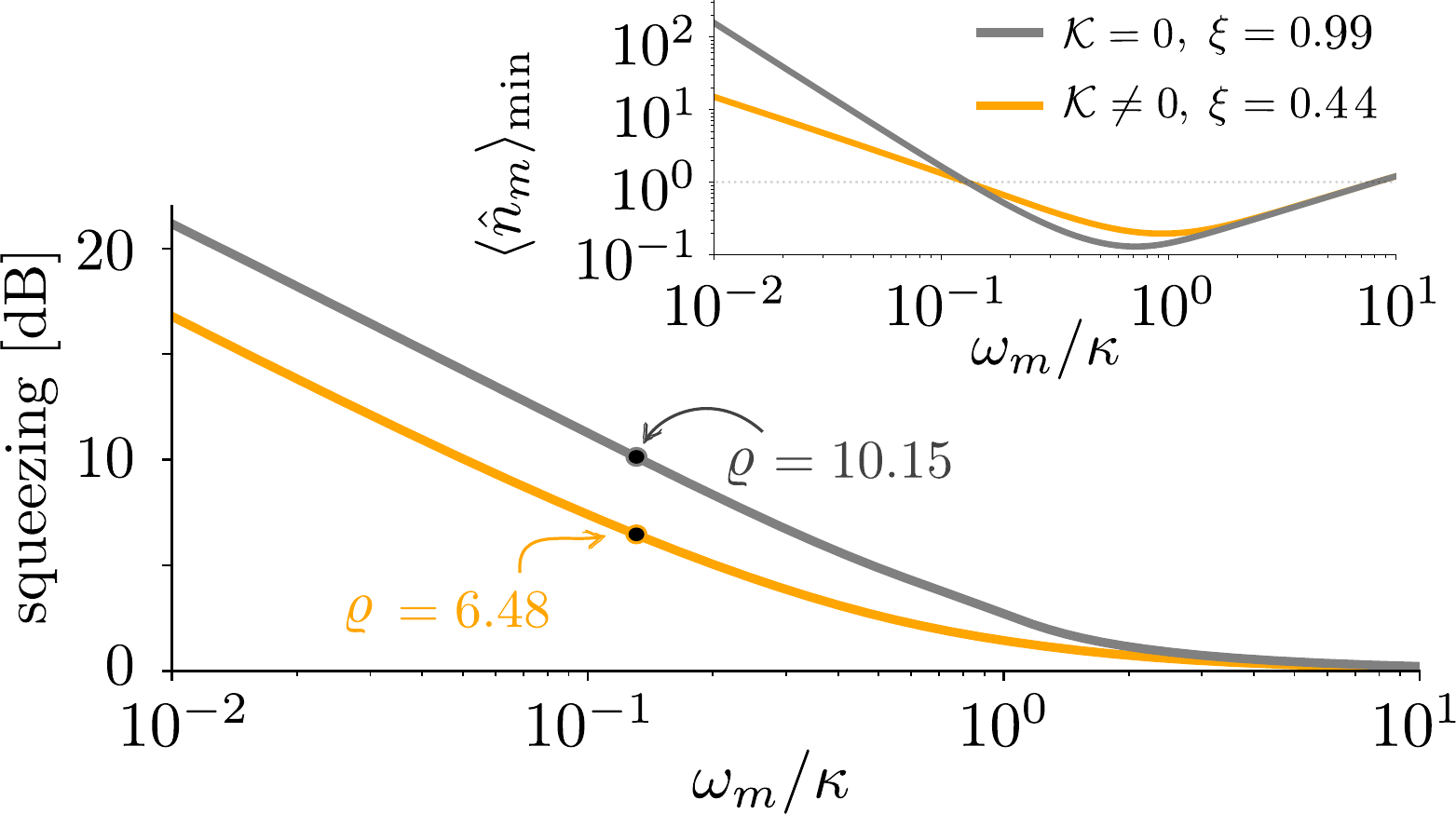}
	\caption{Squeezing for optimal cooling as a function of the resolved sideband parameter for
    the nonlinear (orange line) and linear (grey line) case, using a purity of squeezing of $\xi = 0.44$ and $\xi = 0.99$, respectively.
    The two black dots denote the respective squeezing strengths at $\omega_m/\kappa = 0.13$ discussed in the main text.  
    The inset shows the corresponding minimum phonon number as a function of the resolved parameter. Here, mechanical occupation below one is obtained at $\omega_m/\kappa = 0.13$, where both lines intersect. For the plots we used $g_0/2\pi = 15 \text{ kHz}$ and remaining parameters as in Table~\ref{table_param}.}
	\label{fig:Squeezing_factor}
\end{figure}

Alternatively one can ask the question on the squeezing requirements to achieve a mechanical occupation below one in the linear versus nonlinear case. To discuss this we fix the coupling strength to $g_0/2\pi = 15 \text{ kHz}$ and choose the resolved sideband parameter as $\omega_m/\kappa = 0.13$.  
As indicated by the two black dots in Fig.\ \ref{fig:Squeezing_factor} we find that by using squeezed vacuum with $\xi = 0.99$ for the linear cavity, the unwanted cavity backaction is sufficiently suppressed, allowing the linear case to reach an occupation below one at $\omega_m/\kappa = 0.13$. 
We can connect the required amount of squeezing via $\xi N_s \equiv \sinh^2 (r)$ to an effective \textit{squeezing strength}  
of $\varrho 
= 10 \log e^{2r} =  10.15\; \text{dB}$ to achieve these results (see App. \ref{app_squeezing}).  
In contrast, under the same parameters, a nonlinear cavity requires lower purity ($\xi = 0.44$) and thus lower squeezing strength ($\varrho = 6.48\; \text{dB}$) than a linear cavity, to achieve equivalent cooling limits. This highlights the superior cooling efficiency of a nonlinear cavity compared to a linear system in the unresolved sideband regime, as visible in Fig.\ \ref{fig:Squeezing_factor} which depicts the requirements on the squeezing strength as a function of the resolved sideband parameter. To note is, that for increasing the sideband resolved parameter, best cooling arises when $\Delta_\text{eff} \approx \omega_m$. From Eq.\ \eqref{wp_value} the above results into $\wp^2 \to 0$ leading to a vanishing squeezing strength.  
Additionally, when approaching the resolved sideband regime the cooling benefits of the nonlinear cavity diminish. This regime is also not limited by the backaction of the cavity, hence the squeezing protocol looses its benefit.  

\section{Conclusions}\label{sec_conclusions}
In summary, in this work we showed that a nonlinear cavity, which is dispersivly coupled to a low-frequency mechanical oscillator, is more efficient in dynamical backaction cooling compared to an identical linear system. More efficient means, that for achieving low occupations
in the unresolved sideband regime, the requirements on optomechanical coupling strengths and driving power are significantly lower in the nonlinear case.

We illustrated that the nonlinear nature of the cavity is crucial in the description of the optomechanical backaction. In this regard, the description of the classical dynamics is vital in this study, since the strongest cooling appears at the point of bifurcation of the classical cavity dynamics. We showed that this enhanced cooling performance results from the distinctive asymmetrical photon number spectrum of the nonlinear cavity.
Moreover, we found that the limits for dynamical backaction cooling coincide in the linear and nonlinear case. However, when driven with squeezed vacuum to mitigate the unwanted backaction heating, a nonlinear cavity demonstrates again superior cooling efficiency. For identical parameters, ground state cooling with a nonlinear cavity demands considerably less squeezing strength than a linear system, which is very intriguing from an experimental perspective.

Despite being considered separately as unwanted features, the combination of an optomechanical system in the unresolved sideband regime with an intrinsically nonlinear cavity leads to a better cooling performance. This makes the nonlinear cavity an appealing system for applications involving large and thus low-frequency mechanical systems.

\section{Acknowledgements}
\label{sec_thanks}
%
We want to thank Hans Huebl, John Teufel and Witlef Wieczorek for the fruitful discussions. N.D.N and A.M. acknowledge funding by the project CRC 183. A.M. acknowledges funding by the Deutsche Forschungsgemeinschaft through the Emmy Noether program (Grant No.~ME 4863/1-1). D. Z. is funded by the European Union’s Horizon 2020 research and innovation program under grant agreements no. 736943 and no. 101080143. M. L. J. acknowledges funding from the NSERC and the Canada First Research Excellence Fund. C. M. F. S. and L. D. are supported by the Austrian Science Fund FWF within the DK-ALM (W1259-N27). 
This work was supported by the European Union‘s Horizon Europe
2021-2027 Framework Programme under Grant Agreement No. 101080143 (SuperMeQ).

\appendix

\section{Derivation effective Hamiltonian}
The dynamics of the system can be analyzed within the framework of the master equation, in which the cavity is considered as a reservoir responsible for inducing both heating and cooling processes in the mechanical mode, specifically referred to as Stokes and anti-Stokes processes, respectively. The master equation reads
\begin{align}
	\begin{split}
		\frac{d}{dt}\hat{\rho} &= - i \left[ \hat{\mathcal{H}}, \hat{\rho} \right] + \kappa \hat{\mathcal{D}}\left[ \hat{a} \right] \hat{\rho}
		\\
		&+ \gamma_m \left( \bar{n}_m^T + 1 \right)\hat{\mathcal{D}}\left[ \hat{b} \right] \hat{\rho}
		+ \gamma_m \bar{n}_m^T \hat{\mathcal{D}}\left[ \hat{b}^\dagger \right] \hat{\rho},
	\end{split}
	\label{master_composite}
\end{align}
where $\hat{\rho} \in \mathcal{H}_c \otimes \mathcal{H}_m$ is the density operator of the composite system with $\mathcal{H}_{c(m)}$ the Hilbert space of the optical(mechanical) mode, $\hat{\mathcal{H}}$ is the Hamiltonian in Eq.\ \eqref{Hamiltonian_FirstApprox} and $\hat{\mathcal{D}}[\hat{o}] \bullet = \hat{o} \bullet \hat{o}^\dagger - \{\hat{o}^\dagger \hat{o}, \bullet\}/2$ with $\bullet$ a place-holder.

Recalling the displacement operator for the cavity mode $\hat{\mathcal{D}}(\alpha) = \exp\{\alpha \hat{a}^\dagger - \alpha^* \hat{a}\}$ and the Baker-Hausdorff formula \cite{sakurai2017modern} we find that $\hat{\mathcal{D}}^\dagger(\alpha) \hat{a} \hat{\mathcal{D}}(\alpha)  = \alpha + \hat{d}$, with the time-dependent classical amplitude $\alpha = \alpha(t)$. Similarly, we introduce the displacement operator for the mechanical $\hat{\mathcal{D}}^\dagger(\beta)$, whose action yields $\hat{\mathcal{D}}^\dagger(\beta) \hat{b} \hat{\mathcal{D}}(\beta)  = \beta + \hat{b}$, with the time-dependent classical amplitude $\beta = \beta(t)$ and where, for simplicity, we kept the operator $\hat{b}$ to describe the fluctuations in the mechanics. 

We now continue transforming Eq.\ \eqref{master_composite} to a displaced frame $\hat{\rho}  \to \hat{\rho}' = \hat{P} \hat{\rho} \hat{P}^\dagger$ with the unitary $\hat{P}(t) = \hat{\mathcal{D}}^\dagger(\alpha) \hat{\mathcal{D}}^\dagger(\beta)$ which results in
\begin{align}
	\begin{split}
		\frac{d}{dt}\hat{\rho}' &= - i \left[ \hat{\mathcal{H}}', \hat{\rho}' \right] + \kappa \hat{\mathcal{D}}\left[ \alpha + \hat{d} \right] \hat{\rho}'
		\\
		&+ \gamma_m \left( \bar{n}_m^T + 1 \right)\hat{\mathcal{D}}\left[ \beta + \hat{b} \right] \hat{\rho}'
		+ \gamma_m \bar{n}_m^T \hat{\mathcal{D}}\left[ \beta^* + \hat{b}^\dagger \right] \hat{\rho}',
	\end{split}
	\label{master_transformed_displaced}
\end{align} 
with the transformed Hamiltonian
\begin{align}
	\hat{\mathcal{H}}' = \hat{P}(t) \hat{\mathcal{H}} \hat{P}^\dagger(t) + i\frac{\partial \hat{P}(t)}{\partial t} \hat{P}^\dagger(t).
	\label{Hamiltonian_transformed}
\end{align}
Here, the second term does not vanish due to the time-dependence of the classical mode amplitudes and using the product rule it follows that  
\begin{align}
	\begin{split}
		i\frac{\partial \hat{P}(t)}{\partial t} \hat{P}^\dagger(t) = -i \left[ \frac{\partial \alpha}{\partial t}  \hat{d}^\dagger - \frac{\partial \alpha^*}{\partial t} \hat{d} \right] -i \left[ \frac{\partial \beta}{\partial t}  \hat{b}^\dagger - \frac{\partial \beta^*}{\partial t} \hat{b} \right].
	\end{split}
	\label{Hamiltonian_transformed2}
\end{align}

To find the transformed Hamiltonian in Eq.\ \eqref{Hamiltonian_transformed} we recall the analysis done in Sec. \ref{sec_classical_dynamics} yielding the explicit expression for the dynamics of the cavity amplitude 
\begin{align}
	\frac{d}{dt} \alpha = \left(i\Delta - \frac{\kappa}{2} \right) \alpha + i \mathcal{K} |\alpha|^2 \alpha - i \frac{g_0}{x_\text{zpf}} \langle \hat{x} \rangle \alpha  
	- i \alpha_p
	\label{dynamic_alpha}
\end{align}
$\langle \hat{x} \rangle = x_\text{zpf} \langle \hat{b} + \hat{b}^\dagger \rangle$. Analogously, for the mechanics we find
\begin{align}
	\frac{d \beta}{d t} = - \left(i \omega_m + \frac{\gamma_m}{2} \right) \beta -i g_0 |\alpha|^2.
	\label{dynamic_beta}
\end{align} 
The substitution of Eqs.\ \eqref{dynamic_alpha} and \eqref{dynamic_beta} into Eq.\ \eqref{Hamiltonian_transformed} using Eq.~\eqref{Hamiltonian_transformed2} leads us to the following effective description of the master equation 
\begin{align}
	\begin{split}
		\frac{d}{dt}\hat{\rho}' &= - i \left[ \mathcal{H}', \hat{\rho}' \right] + \kappa \hat{\mathcal{D}}\left[ \hat{d} \right] \hat{\rho}'
		\\
		&+ \gamma_m \left( \bar{n}_m^T + 1 \right)\hat{\mathcal{D}}\left[ \hat{b} \right] \hat{\rho}'
		+ \gamma_m \bar{n}_m^T \hat{\mathcal{D}}\left[ \hat{b}^\dagger \right] \hat{\rho}',
	\end{split}
	\label{master_transformed_displaced_explicit}
\end{align} 
with the modified Hamiltonian
\begin{align}
	\begin{split}
		\hat{\mathcal{H}}' =& 
		-\left(\Delta + 2\mathcal{K} |\alpha|^2 - \frac{g_0}{x_\text{zpf}} \langle \hat{x} \rangle \right) \hat{d}^\dagger \hat{d} 
		+ \omega_m \hat{b}^\dagger \hat{b}
		\\
		&- \frac{\mathcal{K}}{2} \left( \alpha^2  \hat{d}^{2\dagger} + \alpha^{*2}  \hat{d}^2 \right)  
		+ g_0\left[ \alpha^* \hat{d} + \alpha \hat{d}^\dagger\right] \left[  \hat{b} + \hat{b}^\dagger \right] \\
		&+ \hat{\mathcal{H}}_\text{NL},
	\end{split} 
\end{align}
where $\hat{\mathcal{H}}_\text{NL}$ contains all non-quadratic terms, which we will neglect.

Furthermore, in the weak coupling regime $g_0 \ll \kappa$ we have that $g_0 \langle \hat{x} \rangle/x_\text{zpf}$ is small, such that the effective cavity detuning can be denoted $\Tilde{\Delta}= \Delta + 2\mathcal{K} |\alpha|^2$. Finally, the effective Hamiltonian obtained from the unitary transformation of Eq.\ \eqref{master_composite} is given by Eq.\ \eqref{Hamiltonian_Rotated}.

\section{Cavity bistability} \label{sec_explicit_bistable}
We are interested in driving strengths that lead to a large average number of intracavity photons $\bar{n}_c$, but do not cause multistable solutions in the classical cavity dynamics. Following \cite{input_output} the dynamics of the cavity average amplitude reads
\begin{align}
	\frac{d}{dt} \alpha =  \left( i \Delta - \frac{\kappa}{2} \right) \alpha + i \mathcal{K} |\alpha|^2 \alpha - i \sqrt{2} g_0 \langle \hat{q} \rangle  \alpha - \sqrt{\kappa} \alpha_\text{in},
	\label{class_cav}
\end{align}
where $\hat{q} = ( \hat{b} + \hat{b}^\dagger)/ \sqrt{2}$ and $\alpha_\text{in}$ are the mechanical displacement quadrature operator and the coherent drive amplitude, respectively. Analogously, the classical dynamics of the mechanical mode are given by
\begin{align}
	\begin{split}
		\frac{d}{dt} \langle \hat{q} \rangle &= \omega_m \langle \hat{p} \rangle - \frac{\gamma_m}{2} \langle \hat{q} \rangle,
		\\
		\frac{d}{dt} \langle \hat{p} \rangle &= -\omega_m \langle \hat{q} \rangle - \frac{\gamma_m}{2} \langle \hat{p} \rangle - \sqrt{2} g_0 |\alpha|^2
	\end{split}
	\label{mech_eq}
\end{align}
with $\hat{p} = i( \hat{b}^\dagger - \hat{b})/ \sqrt{2}$ and $\gamma_m$ the momentum quadrature operator and the decay rate of the mechanical oscillator, respectively.

Since the oscillation of the mechanical position is small enough to weakly modulate the cavity field, we solve Eq.\ \eqref{mech_eq} in the long time limit and find the steady state of the mechanical position operator
\begin{align}
	\langle \hat{q} \rangle_s = - \frac{\sqrt{2} g_0 \omega_m |\alpha|^2}{\omega^2_m + \frac{\gamma_m^2}{4}},
\end{align}
which we insert into the equation for the classical cavity amplitude given in Eq.\ \eqref{class_cav} and obtain
\begin{align}
	\frac{d}{dt} \alpha = \left( i \Delta - \frac{\kappa}{2} \right) \alpha + i \mathcal{K}_\text{eff} \alpha |\alpha|^2 - \sqrt{\kappa} \alpha_\text{in},
	\label{steady_sol_cav}
\end{align}
with the effective Kerr constant given in Eq.\ \eqref{effective_kerr}. The multiplication of the steady state solution of Eq.\ \eqref{steady_sol_cav} with its complex conjugate yields ultimately the average photon occupation in Eq.\ \eqref{cubic_photon}. The possible values of the average cavity number can be explicitly calculated yielding 
\begin{align}
	|\alpha|_1^2 &= \frac{1}{3 \mathcal{K}_\text{eff}} \left\{ -2 \Delta - \Sigma + \frac{\Lambda_0}{\Sigma} \right\}, \label{Class_Sol_1}\\
	|\alpha|_{2,3}^2 &= \frac{1}{3\mathcal{K}_\text{eff}}
	\left\{ - 2 \Delta + e^{\mp i \frac{\pi}{3}} \Sigma - e^{\pm i \frac{\pi}{3}} \frac{\Lambda_0}{
		\Sigma} \right\} \label{Class_Sol_2}
\end{align}
where we introduced the following definitions $\Sigma = \sqrt[3]{\sqrt{\Lambda_0^3 + \Lambda_1^2} + \Lambda_1}$ with $\Lambda_0 = \frac{3\kappa^2}{4}- \Delta^2$ and $\Lambda_1 = -\left(\frac{9\kappa^2}{4} + \Delta^2 \right) \Delta - \frac{27}{2} \kappa\, \mathcal{K}_\text{eff} \bar{n}_\text{in}$.

\section{Derivation of the radiation pressure force spectrum} \label{app_force_spectra}
The radiation pressure force spectrum  describes the strength of the fluctuations of the cavity photon number at different frequencies. In the regime of linearized optomechanics this spectrum is obtained via 
\begin{align}
	\begin{split}
	    S_{FF}[\omega] = \int_{-\infty}^\infty d\omega' \langle \hat{F}[\omega] \hat{F}[\omega'] \rangle  = g_0^2 S_{nn}[\omega],
	\end{split}
\end{align}
with the linearized radiation pressure force is simply given by $\hat{F}[\omega] = G^*\hat{d}[\omega] + G\hat{d}^\dagger[\omega]$. For simplicity, here we will rewrite the solution obtained in Eq.\ \eqref{kerr_cavity_fluctuations} as 
\begin{align}
    \hat{d}[\omega] = - \frac{\sqrt{\kappa}}{\mathcal{N}[\omega]}
    \left\{ \mathcal{X}^{*-1}_c[-\omega] \hat{d}_\text{in} + i \Lambda \hat{d}_\text{in}^\dagger \right\},
\end{align}
with the definition $\mathcal{N}[\omega] = \mathcal{X}_c^{*-1}[-\omega] \mathcal{X}_c^{-1}[\omega] - |\Lambda|^2$. This allows us to express the linearized radiation pressure force as 
\begin{align}
	\hat{F}[\omega] = \eta^*[-\omega] \hat{d}_\text{in}[\omega] + \eta[\omega] \hat{d}_\text{in}^\dagger[\omega]
 \label{force_operator}
\end{align}
with 
\begin{align}
\eta[\omega] = -\frac{\sqrt{\kappa}}{\mathcal{N}[\omega]} \left( G\mathcal{X}_{c}^{-1}[\omega] + i G^* \Lambda \right).
\end{align}

Assuming delta correlated noise as in Eq.\ \eqref{White_noise_correl} we find that the only non-zero contribution comes from the correlator  $\langle \hat{d}_\text{in}[\omega']\hat{d}^\dagger_\text{in}[\omega] \rangle$, such that the spectrum of the radiation pressure force reads
\begin{align}
\begin{split}
    \mathcal{S}_{FF}[\omega] &=  \int \frac{d \omega'}{2 \pi}\, \eta^*[-\omega]\eta[\omega'] 
    \,\langle \hat{d}_\text{in}[\omega]\hat{d}^\dagger_\text{in}[\Omega'] \rangle,  
    \label{force_spectra_vacuum}
\end{split}
\end{align}
which coincides with Eq.\ \eqref{stokes_anti_stokes} with the photon number spectrum given by Eq.\ \eqref{photon_spectra}.

\section{Derivation of the average mechanical occupation} \label{sec_derivation_mech_occ}
The occupation of the mechanical mode can be obtained using Eq.\ \eqref{mech_occupation_spectra} with the noise spectrum given by Eq.\ \eqref{spectra_quality_factor} and which can be calculated through the following equation
\begin{align}
	\mathcal{S}_{bb}[\omega] = \int \frac{d \Omega}{2\pi} 
	\langle \hat{b}^\dagger [\Omega] \hat{b}[\omega] \rangle,
	\label{app_mech_spectra}
\end{align}
with $\hat{b}[\omega]$ given by Eq.\ \eqref{Eff_Quantum_Langevin_Mechanics}, such that  within the high-Q approximation last equation becomes Eq.\ \eqref{spectra_quality_factor}, whose poles $\mathcal{P}_\pm$ become
\begin{align}
	\Omega_\pm  = \pm \left[ \mathcal{W}_R - i \mathcal{Z}_\pm \right],
	\label{poles_app}
\end{align}
where we defined $\mathcal{W}_R - i \mathcal{W}_I \equiv \sqrt{\omega_m^2 - 2 \omega_m \Sigma_c[\omega_m]}$ and $\mathcal{Z}_\pm \equiv \pm \gamma_m/2 + \mathcal{W}_I$. 

To perform the integration of Eq.\ \eqref{spectra_quality_factor} via the Residue theorem \cite{Lang1999}, we have to first perform a stability analysis of the poles. Here, the only possible stable solution arises when $\mathcal{Z}_\pm > 0$. Moreover, under the assumption that  $\Sigma_c[\omega_m]/\omega_m \ll 1$ we can make a Taylor expansion and find that $\mathcal{W}_R = \omega_m - \Re \left\{ \Sigma_c[\omega_m]\right\}$ and $\mathcal{W}_I =  \Im \left\{ \Sigma_c[\omega_m]\right\}$. This is equivalent to neglecting the off-diagonal entries in the effective susceptibility matrix given by Eq.\ \eqref{susceptibilit_effective}, which leads to off-resonant processes. Based on the previous analysis and using the Residue theorem, we find the following integrals
\begin{align}
	\begin{split}
		&\int_{-\infty}^\infty \frac{d\omega}{2\pi}\;\frac{1}{\left|\left(\omega - \Omega_+\right)\left(\omega -  \Omega_-\right)\right|^2} 
		\approx \frac{\mathcal{C}_\text{eff}}{2\omega_m^2 \gamma_m\left[  \mathcal{C}_\text{eff}^2 - 1 \right]},
		\\
		&\int_{-\infty}^\infty \frac{d\omega}{2\pi}\; \frac{\omega}{\left|\left(\omega - \Omega_+\right)\left(\omega - \Omega_-\right)\right|^2}
		\approx  \frac{1}{2\omega_m	 \gamma_m\left[ 1 - \mathcal{C}_\text{eff}^2 \right]},
		\\
		&\int_{-\infty}^\infty \frac{d\omega}{2\pi}\; \frac{\omega^2}{\left|\left(\omega - \Omega_+\right)\left(\omega - \Omega_-\right)\right|^2}
		\approx \frac{\mathcal{C}_\text{eff} \left(1  - \left(\frac{\gamma_m}{2 \omega_m}\right)^2\right)}{ 2\gamma_m\left[  \mathcal{C}_\text{eff}^2 - 1 \right]}
	\end{split}
	\label{integrals}
\end{align}
from which we obtain the average mechanical occupation given in Eq.\ \eqref{mech_occ_steady}.
\section{Mechanical occupation: master equation approach} \label{sec_equivalence}
If we work in the framework of an effective master equation for the mechanical mode we find that the average mechanical occupation follows Eq.\ \eqref{mech_occ_steady_state} with the induced damping given by $\Gamma_\text{opt} = \Gamma_\text{AS} - \Gamma_\text{S}$, such that cooling is only possible if $\Gamma_\text{AS} > \Gamma_\text{S}$. The first term in Eq.\ \eqref{mech_occ_steady_state} is effectively a modified mechanical thermal occupation, whereas the second term is what we know as the cavity backaction limit. 

Recalling that the cooperativity can be written in terms of the induced damping, since $\Gamma_\text{opt} = 2 \Im\{\Sigma_c[\omega_m]\}$ and consequently $2 \mathcal{C}_\text{eff} = \Gamma_\text{opt}/\gamma_m$, we find that Eq.\ \eqref{mech_occ_steady_state} becomes
\begin{align}
	\bar{n}_m \approx \frac{\bar{n}_m^T}{1 + 2 \mathcal{C}_\text{eff}} + \left( \frac{  2 \mathcal{C}_\text{eff} }{1 + 2 \mathcal{C}_\text{eff}} \right)\frac{\Gamma_S}{\Gamma_\text{opt}},
	\label{phonon_number}
\end{align}
where the explicit substitution of the Stokes and anti-Stokes rates Eq.\ \eqref{stokes_anti_stokes} results in Eq.\ \eqref{mech_occ_steady}.

\begin{figure*}[ht]
    \centering 
    \includegraphics[width=\textwidth]{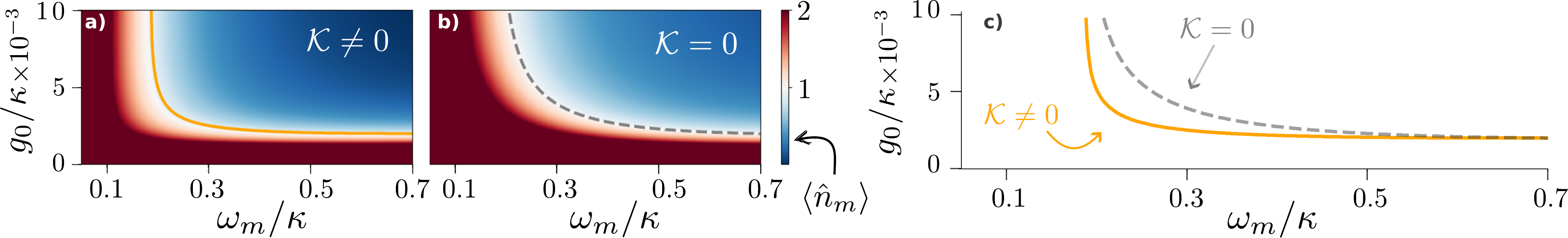}
    \caption{Average phonon number as a function of the optomechanical coupling strength $g_0/\kappa$ and resolved sideband parameter $\omega_m/\kappa$ using a nonlinear a) and linear system b). c) The solid orange and dashed grey lines show the values for which the ground state is achieved in a nonlinear and linear setup, respectively. Ground state cooling can be obtained using a nonlinear cavity  both at lower coupling strength and smaller mechanical frequency as for an equivalent linear system.. All other parameters are given in Table \ref{table_param}.}
	\label{fig:Ground_state}
\end{figure*}

\section{Minimal occupation}
\label{app_minOcc}
We can combine our discussion in the main text and vary both parameters, the coupling strength and the resolved sideband parameter while keeping the input power at $\bar{n}_\text{in} = \bar{n}_\text{in,crit}$.
In Fig.~\ref{fig:Ground_state} we show when ground state cooling is achieved as a function of the optomechanical coupling strength $g_0/\kappa$ and the resolved sideband parameter $\omega_m/\kappa$. At $g_0/\kappa = 0.85 \times 10^{-3}$, a nonlinear cavity (orange line) can cool to the ground state when $\omega_m/\kappa = 0.189$. Conversely, for the same input power
a linear system requires both a larger coupling strength of $g_0/\kappa = 1.79 \times 10^{-3}$ and a slightly larger mechanical frequency $\omega_m/\kappa = 0.192$. 
Hence, we find that the requirement for the resolved sideband parameter to be at $\omega_m/\kappa > 0.177 \approx 0.2$ is a good approximation 
for ground state cooling to be feasible, and that it is necessary in both cases. However, for the same input power the linear cavity requires roughly twice the coupling strength ratio $g_0/\kappa$.
This again shows that a nonlinear cavity can cool at lower coupling strength in the unresolved sideband regime. 

\section{Suppression of the Stokes process} \label{app_squeezing}
In Sec. \ref{sec_cooling}, we demonstrated that in the unresolved sideband regime, a nonlinear cavity outperforms a linear system. Nevertheless, despite this improvement, cooling to the ground state in this regime is still constrained by the cavity backaction given explicitly in Eq.\ \eqref{cavity_backaction}.   Consequently, to overcome this limitation and cool below the backaction limit we inject  squeezed vacuum, which can be externally generated by a degenerate parametric amplifier (DPA) \cite{Carmichael_1984_DPA}. The dynamics of the DPA is described by the quantum Langevin equation
\begin{align}
    \dot{c} = - \frac{\zeta}{2} \hat{c} + \chi \hat{c}^\dagger - \sqrt{\zeta} \hat{c}_\text{in}
    \label{DPA_equation}
\end{align}
with $\sqrt{\zeta} \hat{c}_\text{in} = \sqrt{\zeta_\text{s}} \hat{c}_\text{in,s} + \sqrt{\zeta_\text{v}} \hat{c}_\text{in,v}$ the input field; $\zeta = \zeta_\text{s} + \zeta_\text{v}$ the total decay rate, and $\chi = |\chi| e^{-2i \varphi}$ is the squeezing parameter with $\varphi$ the squeezing angle. Here, the input noise $\hat{c}_\text{in,v}$ corresponds to the intrinsic losses of the DPA, whose corresponding rate is given by $\zeta_\text{v}$. On the other hand, $\hat{c}_\text{in,s}$ is related to the input/output port of the DPA, i.e., to the external modes of the electromagnetic field, that are controlled and used to drive the optomechanical system.
The output field of the DPA is cascaded into the optomechanical system yielding new cavity noise operators, which can be written as $\sqrt{\kappa} \hat{d}_\text{in} = \sqrt{\kappa_\text{s}} \hat{c}_\text{out} + \sqrt{\kappa_\text{v}} \hat{d}_\text{in,v}$ with $\kappa = \kappa_\text{s} + \kappa_\text{v}$ and $\hat{c}_\text{out} = \sqrt{\zeta_\text{s}} \hat{c} + \hat{c}_\text{in,s}$. Here, $\hat{c}_\text{out}$ is the output operator of the squeezed bath, which exchanges photons with the cavity at rate $\kappa_\text{s}$, and $\hat{d}_\text{in,v}$ is related to the uncontrolled losses of the cavity at rate $\kappa_\text{v}$. To quantify the combined intrinsic losses in the DPA and the nonlinear cavity we define the loss parameter $\xi = \kappa_{s} \zeta_{s}/ (\kappa \zeta)$.
Using the \textit{white-noise approximation} \cite{Gardiner_Noise} the new noise correlators become Eq.\ \eqref{White_noise} with the definitions Eq.\ \eqref{White_noise_correl}. 

Similarly as in Sec. \ref{sec_radiation_pressure_force}, using the noise correlators in Eq.\ \eqref{White_noise}, we derive the radiation pressure force spectrum  
\begin{align}
	\begin{split}
		&\mathcal{S}_{FF}[\omega]
		=
		\mathcal{S}_{FF}^0[\omega]
		\left\{ 1 + \left[ 1 + \frac{\left[\omega - \Delta_\text{eff}  \right]^2 + \frac{\kappa^2}{4}}{\left[\omega + \Delta_\text{eff}  \right]^2 + \frac{\kappa^2}{4}} \right]\xi N_s\right\} 
		\\
		&- \frac{2 \xi M_s\left\{ \left( \Delta_\text{eff}^2 - \omega^2 - \frac{\kappa^2}{4} \right)   \cos(2 \varphi) - \Delta_\text{eff} \kappa \sin(2 \varphi) \right\} }{\left[\tilde{\Delta}^2 - \omega^2 + \frac{\kappa^2}{4} - |\Lambda|^2  \right]^2 + \kappa^2 \omega^2},
	\end{split}
	\label{Force_spectra_squeezed}
\end{align}
which in the absence of squeezing, i.e. $\xi = 0$, we recover $\mathcal{S}_{FF}^0[\omega] = g_0^2 \mathcal{S}^0_{nn}[\omega]$ with the photon number spectrum given by Eq.\ \eqref{photon_spectra}. 

To minimize the Stokes process, an extrema analysis was conducted to determine the optimal squeezing phase, which is expressed as follows
\begin{align}
	\varphi  = \frac{1}{2} \arctan \left( \frac{\kappa \Delta_\text{eff} }{\omega_m^2 - \Delta_\text{eff}^2  + \frac{\kappa^2}{4} } \right) + k \pi
    \label{optimal_phase}
\end{align}
with $k \in \mathbb{Z}$. Hence, using the last expression we find that the cavity backaction limit becomes Eq.\ \eqref{Backlimit_Main} with $\bar{n}_\text{BA}$ given by Eq.\ \eqref{cavity_backaction} and Eq.\ \eqref{wp_value}. Note that in the absence of squeezing $\xi = 0$ in Eq.\ \eqref{Backlimit_Main} we recover $\bar{n}_\text{BA}$, which is the backaction corresponding to a cavity with solely input vacuum noise. 
We can further minimize the backaction limit Eq.\ \eqref{Backlimit_Main} over the properties of the squeezed input light and find that for $\wp = \sqrt{N_s/(N_s + 1)}$ we obtain Eq.\ \eqref{squeezed_BA}. 
As expected, in the absence of squeezing we recover Eq.\ \eqref{cavity_backaction}, but for $\xi = 1$  we can fully suppress the quantum backaction limit.

In addition, we can quantify the amount of squeezing using the \textit{squeezing factor} $r$, which is related to the noise correlators in Eq.\ \eqref{White_noise_correl} through the following relation
\begin{align}
    N_s = \frac{1}{\xi} \sinh^2(r), \quad M_s = \frac{1}{\xi} \sinh(r) \cosh(r).
\end{align}
Thus, from the condition yielding Eq.\ \eqref{squeezed_BA}, we find that the squeezing factor is given by the following equation
\begin{align}
    \sinh^2(r) = \frac{\xi \wp^2}{1 - \wp^2} \ge 0
    \label{squeezing_strength}
\end{align}
with $\wp$ given by Eq.\ \eqref{wp_value}. Finally, in Fig.\ \ref{fig:Minima_Resolved_squeezing} we measured the squeezing in units of dB using the formula
  $  \varrho = -10 \log_{10} \left(e^{-2r} \right)$.



\end{document}